\documentclass[twocolumn]
{aastex631}
\usepackage{graphicx}
\usepackage[normalem]{ulem}
\usepackage{amsmath}
\usepackage{dsfont}

\def\cos{\text{cos}}
\def\sin{\text{sin}}
\def\lp{\left(}
\def\rp{\right)}
\def\lb{\left[}
\def\rb{\right]}
\def\lc{\left\{}
\def\rc{\right\}}
\def\t{\text}

\usepackage{color}
\usepackage{here}
\usepackage{float}
\usepackage{tensor}
\usepackage{xspace}
\usepackage{orcidlink}
\usepackage{url}
\usepackage[multiple]{footmisc}

\def\Mc{M_{\rm cloud}} 
\def\Rc{R_{\rm cloud}} 
\def\rhoc{\rho_{\rm cloud}}
\def\tff{t_{\rm ff}}
\def\nocrs{\rm M2e3\_noCRT}
\def\fiducial{\rm M2e3\_E1\_D8e25}
\def\highcrs{\rm M2e3\_E10\_D8e25}
\def\lowdiffcoeff{\rm M2e3\_E1\_D8e24}
\def\highdiffcoeff{\rm M2e3\_E1\_D8e26}
\def\vardiffcoeff{\rm M2e3\_E1\_DVAR}

\def\vcr{v_{\rm cr}}

\def\gammacr{\gamma_{\rm cr}}

\def\ecr{e_{\rm cr}} 
\def\Fecr{F_{e, \rm cr}}
\def\Pcr{P_{\rm cr}}
\def\epscr{\epsilon_{\rm cr}}
\def\Lambdacr{\Lambda_{\rm cr}}

\def\bhat{\hat{\bold{B}}}

\def\Dop{\mathds{D_{\rm cr}}}

\def\Pop{\mathds{P}_{\rm cr}}

\def\Ssc{\tilde{S}_{\rm sc}}

\def\Se{S_{e}}
\def\Sfe{S_{F_{e}}}
\def\Gammacr{\Gamma_{\rm cr}}
\def\Gammast{\Gamma_{\rm st}}

\def\barnu{\bar{\nu}_s}

\def\muftwo{\langle\mu_f^2\rangle}
\def\mufone{\langle\mu_f^1\rangle}

\def\vgas{\bold{u}} 

\def\Sgas{S_{\rm gas}}

\def\vst{v_{\rm st}}
\def\vAi{v_{\rm Ai}}

\newcommand{\cut}[1]{{\textcolor{red}{\sout{#1}}}}

\shorttitle{Short Title}
\shortauthors{Fitz Axen et al.}

\begin{document}

\title{Suppressed Cosmic Ray Energy Densities in Molecular Clouds From Streaming Instability-Regulated Transport}

\correspondingauthor{Margot Fitz Axen}
\email{fitza003@utexas.edu}

\author[0000-0001-7220-5193]{Margot Fitz Axen}
\affiliation{Department of Astronomy, The University of Texas at Austin, 2515 Speedway, Stop C1400, Austin, Texas 78712-1205, USA.}

\author[0000-0001-7220-5193]{Stella Offner}
\affiliation{Department of Astronomy, The University of Texas at Austin, 2515 Speedway, Stop C1400, Austin, Texas 78712-1205, USA.}

\author[0000-0003-3729-1684]{Philip F. Hopkins}
\affiliation{TAPIR, Mailcode 350-17, California Institute of Technology, Pasadena, CA 91125, USA}

\author[0000-0003-3893-854X]{Mark R. Krumholz}
\affiliation{Research School of Astronomy and Astrophysics, Australian National University, Canberra ACT 2600, Australia}
\affiliation{ARC Center of Excellence for Astronomy in Three Dimensions (ASTRO-3D), Canberra ACT 2600, Australia}

\author[0000-0002-1655-5604]{Michael Y. Grudi\'{c}}
\affiliation{Carnegie Observatories, 813 Santa Barbara St, Pasadena, CA 91101, USA}

\begin{abstract}
Cosmic rays (CRs) are the primary driver of ionization in star forming molecular clouds (MCs). Despite their potential impacts on gas dynamics and chemistry, no simulations of star cluster formation following the creation of individual stars have included explicit cosmic ray transport (CRT) to date. We conduct the first numerical simulations following the collapse of a $2000 M_{\odot}$ MC and the subsequent star formation including CRT using the STARFORGE framework implemented in the {\small GIZMO} code. We show that when CR-transport is streaming-dominated, the CR energy in the cloud is strongly attenuated due to energy losses from the streaming instability. Consequently, in a Milky Way like environment the median CR ionization rate (CRIR) in the cloud is low ($ \zeta \lesssim 2 \times 10^{-19} \rm s^{-1}$) during the main star forming epoch of the calculation and the impact of CRs on the star formation in the cloud is limited. However, in high-CR environments, the CR distribution in the cloud is elevated ($\zeta \lesssim 6 \times 10^{-18}$), and the relatively higher CR pressure outside the cloud causes slightly earlier cloud collapse and increases the star formation efficiency (SFE) by $50 \%$ to $\sim 13 \%$. The initial mass function (IMF) is similar in all cases except with possible variations in a high-CR environment. Further studies are needed to explain the range of ionization rates observed in MCs and explore star formation in extreme CR environments. 
\end{abstract}


\section{Introduction}
\label{sec:introduction}

Cosmic rays (CRs) are ubiquitous and responsible for driving a variety of astrochemical and dynamic processes. In the interstellar medium (ISM), CRs have an energy density of $\sim$ 1 eV $\rm cm^{-3}$, which is comparable to the energy densities of the cosmic microwave background, starlight, and galactic magnetic field \citep{ferriere_2001}. In the densest regions of molecular clouds (MCs), which are optically thick to UV radiation, CRs are the primary driver of ionization \citep{dalgarno_2006}. The level of CR ionization is usually quantified by the CR primary ionization rate (CRIR), which is the rate of CR ionizations per hydrogen atom per unit time.  

The CRIR in the Solar neighborhood inferred from the CR particle flux measured outside the heliosphere by \textit{Voyager} is $\zeta \approx 1.6 \times 10^{-17} \rm s^{-1}$ \citep{cummings_2016,stone_2019}. Since CR ionizations lead to chemical reactions that produce molecules such as $\t{OH}^+$, $\t{H}_2\t{O}^+$, and $\t{H}_3\t{O}^+$, measurements of these molecules can also be used to constrain the CRIR indirectly, since their abundances are sensitive to it \citep{indriolo_2015}. Observations of these molecules along different sightlines predict a range of values for the CRIR. In the Milky Way, the CRIR varies between $\approx 10^{-17}-10^{-15} \rm s^{-1}$, with values $\approx$ 10 times higher in the Galactic center than in the Galactic disk \citep{indriolo_2015, neufeld_wolfire_2017}. In higher-density regions such as high-mass star-forming regions \citep{sabatini_2020} and low-mass dense cores \citep{caselli_1998, bialy_2022}, the CRIR is expected to be lower due to attenuation by the dense gas. In contrast, recent observations near supernova remnants \citep{ceccarelli_2011} and towards the more active star-forming regions in the center of our Galaxy \citep{rivilla_2022} and NGC 253 \citep{behrens_2022, holdship_2022} suggest orders of magnitude higher CRIRs than would be predicted from \textit{Voyager} data (see review by \citealt{padovani_2023}). 

Some of the observed variation in the CRIR can be explained by the presence of non-SN CR acceleration sources \citep{ceccarelli_2011}. While supernova remnants have long been thought to be the primary source of galactic CRs \citep{blasi_2013}, converting about 10 \% of their energy into CR acceleration \citep{vink_2012}, evidence indicates that stellar winds from massive stars also accelerate CRs \citep{parizot_2004, bykov_2020}. A number of star clusters exhibit $\gamma$-ray emission, indicative of the presence of locally accelerated high-energy CRs \citep{ackermann_2011, yang_aharonian_2017, yang_2018, aharonian_2019, sun_2020, liu_2022, pandey_2024}. Some estimates suggest that stellar winds convert up to $\sim$ 10 \% of their energy into CR acceleration \citep{aharonian_2019}, while others claim it may be as high as 40 \% \citep{pandey_2024}. Most recently, ionization maps of star-forming regions hint that protostellar sources may also accelerate CRs \citep{ceccarelli_2014, cabedo_2023, pineda_2024}, likely accelerated by the accretion or protostellar jet shocks \citep{padovani_2016,gaches_2018}. 

Observations of CRIR variation have motivated a variety of theoretical studies on the impact of CR physics on galactic processes. To date, a growing number of hydrodynamic simulations have included explicit cosmic ray transport (CRT) to study the effect of CRs on galaxy formation and evolution. These calculations suggest that CR pressure may dominate over thermal pressure in the haloes of low redshift Milky Way mass galaxies, driving galactic outflows and suppressing star formation \citep{jubelgas_2008, uhlig_2012, booth_2013, hanasz_2013, salem_2014, pakmor_2016, simpson_2016, wiener_2017, butsky_quinn_2018, jacob_2018, chan_2019, dashyan_dobois_2020, hopkins_2020, ji_2020, farcy_2022, girichidis_2022, girichidis_2023, thomas_2023}. Stratified box calculations of patches of a Milky Way-like disk  that include CRT indicate that including plasma-dependant transport parameters changes the impact of the CRs on the gas structure \citep{girichidis_2016, simpson_2016, farber_2018, holguin_2019}. However, all of these effects strongly depend on the prescription adopted for CRT; for example, \cite{commercon_2019} find that when CRT is treated as purely diffusive, for diffusivities $D \lesssim 10^{25} \rm cm^2/s$ CRs are trapped in the gas by pressure gradients that modify the gas flow. This effect is particularly relevant near CR acceleration sources where the CR diffusivity is suppressed due to turbulence, causing the gas to fragment rather than group into star-forming clumps \citep{semenov_2021}. By contrast, higher diffusivities produce much smaller dynamical effects, because CRs are less effectively trapped and thus large CR pressures do not build up.

Despite evidence from observations and simulations that CR properties vary with environment and change gas properties, most simulations and analytical models of MCs to date  have neglected the impact of CRs altogether \citep[e.g.,][]{padoan_2011, hennebelle_2008, hopkins_2012, li_2015, lee_2019}. Most simulations have assumed a spatially and temporally uniform CRIR of $\approx 10^{-17}-10^{-16} \rm s^{-1}$ to calculate the heating and cooling processes  \citep{glover_clark_2012, bate_keto_2015,gatto_2017, rowan_2020, grudic_2021, guszejnov_2021, guszejnov_2022} or have neglected CR ionization altogether. 
However, in addition to neglecting the impact of CRIR variation on the gas temperature, these simulations do not take into account the possible dynamical effect of CRs on the gas. For example, the level of ionization determines how much the gas couples to the magnetic field and resists collapse \citep{fiedler_1993, padovani_2014}. Simulations of protostellar collapse show that the CR ionization rate also determines 
whether a protostellar disk forms and regulates its size 
\citep{wurster_2018, kueffmeier_2020}. CR energy deposition may also modify the gas turbulence if the CR pressure is high enough  \citep{commercon_2019}. Although some studies have investigated the effects of including CRT in MCs via post-processing, these also neglect the dynamical effects between the CRs and the gas \citep{fatuzzo_adams_2014, fitzaxen_2021}.

No simulations to date have both included CRT and resolved the formation of individual stars. In this study, we carry out large-scale simulations of MC collapse that include CRT in order to follow the propagation of CRs from the galactic CR background. We use the STARFORGE (STAR FORmation in Gaseous Environments) framework, which is built on the {\small GIZMO} simulation code \citep{hopkins_2015} and includes all relevant feedback processes, including protostellar jets, radiative heating, stellar winds, and supernovae. The fiducial STARFORGE MC simulations  
resolve mass scales down to $\approx 0.1 M_\odot$ and so are able to self-consistently predict the stellar initial mass function (IMF) \citep{grudic_2021,grudic_2022,  guszejnov_2021, guszejnov_2022}. In this work we carry out STARFORGE simulations including the {\small GIZMO} single-bin CRT treatment, which has been used extensively in the Feedback in Realistic Environments (FIRE) suite of cosmological simulations \citep{chan_2019, hopkins_2020,hopkins_2021a, hopkins_2021b, hopkins_2021c, hopkins_2021d}. This implementation applied to galaxy-scale calculations is able to reproduce observables such as the CR intensity spectrum observed by \textit{Voyager} \citep{hopkins_2021d}. 

In this work we explore the impact of CRs on ionization, gas temperatures, star formation histories, and the IMF of star-forming MCs. Section \ref{section:methods} describes the physics and numerical methods used in the STARFORGE simulations, the {\small GIZMO} CRT modules, and the simulation initial conditions.
Section \ref{section:results} presents the results of simulations with different recipes for CRT. Section \ref{section:discussion} discusses how CRT properties shape our results and compares to prior work. Finally, Section \ref{section:conclusions} summarizes our results. 

\section{Methods}
\label{section:methods}

\subsection{The STARFORGE Simulations}
\label{subsection:starforge_simulations}

\subsubsection{Physics}
\label{subsubsection:starforge_physics}

We simulate star-forming MCs using the STARFORGE framework, which uses the {\small GIZMO} simulation code \citep{hopkins_2015}. A full description of the STARFORGE methods is included in \citet{grudic_2021}; therefore, we only summarize the key points here. 

We use the Lagrangian constrained-gradient meshless finite-mass method for magnetohydrodynamics (MHD) \citep{hopkins_raives_2016, hopkins_2016} and assume the ideal MHD limit. The simulations utilize the updated FIRE-3 radiation and chemistry modules \citep{hopkins_2023}. They co-evolve the gas, dust, and radiation temperature from the stellar luminosity and an external radiation field.  

Sink particles represent individual stars. Once they form, they follow the subgrid protostellar evolution model in \cite{offner_2009}. The STARFORGE framework includes stellar feedback from protostellar jets, radiation, stellar winds, and supernovae (SNe). Feedback from SNe, radiation, and most stellar winds is injected by distributing mass, momentum, and energy into surrounding gas cells using the weighting scheme described in \cite{hopkins_2018}. Feedback from protostellar jets and stellar winds, where the free-expansion radius is smaller than the size of a wind cell, is injected using the cell spawning technique described in \cite{grudic_2021}. 

Protostellar jets are modeled using the jet feedback model of \cite{cunningham_2011} in which sink particles launch a fraction of their accreted material $f_w$ along their rotational axis with a velocity that is a fraction $f_K$ of the Keplerian velocity defined at the protostellar radius. We use $f_w=f_K=0.3$, which \cite{rosen_krumholz_2020} found reproduces measurements of momentum injection as a function of protostellar luminosity \citep{maud_2015}. We do not include CR acceleration from protostellar jets. 

Massive main-sequence stars inject a significant amount of energy and momentum into their environment through radiative feedback and stellar winds. Winds are launched for stars more massive than $2 M_{\odot}$. The stars lose mass isotropically at a mass-loss rate $\dot{M}_w$ moving at velocity $v_w$. Wind mass-loss rates and velocities are calculated from \citet[][see Equation 1 of \cite{grudic_2022}]{smith_2014} and \citet[][see Equation 45 of \cite{grudic_2021}]{lamers_snow_lindholm_1995}. The luminosity of the winds is $\dot{E}_w=0.5\dot{M}_wv_w^2$. Typical wind luminosities are $\sim 0.1-10 L_{\odot}$ (or $\sim 10^{40}-10^{42}$ erg/yr). Stellar winds with velocities $v_w > 1000 $ km/s inject $10 \%$ of their energy into CRs, consistent with estimates made from $\gamma$-ray measurements in massive star clusters \citep{aharonian_2019, pandey_2024}. The CR energy is injected by including an energy source term $\Se=0.1\dot{E}_w$ and a momentum source term $\Sfe=\Se c(\hat{v_w} \cdot \hat{\bold{B}})$ parallel to the local magnetic field $\hat{\bold{B}}$ in the CR transport equations (Section \ref{subsubsection:cr_transport}). To satisfy conservation of energy the wind velocity is reduced to $v_w \rightarrow \sqrt{0.9}v_w$. 
All of our calculations form sufficiently massive stars such that CRs are injected into the domain.  

The STARFORGE framework  also includes SNe, which are spawned at the end of the lifetime of main sequence stars $> 8 M_{\odot}$. However, all our simulations end before SNe occur. 

\subsubsection{Initial Cloud Properties}
\label{subsubsection:cloud_properties}

Our initial cloud properties are similar to those used in previous STARFORGE simulations \citep{guszejnov_2021, guszejnov_2022, grudic_2021, grudic_2022}. We start with a cloud of gas with mass $\Mc = 2000 M_{\odot}$ and radius $\Rc = 3$ pc at a uniform density of $\rhoc = \Mc/(4\pi\Rc^3/3)$ and a uniform temperature $T_0 = 10$ K. The cloud is placed in a diffuse medium with a density of $\rhoc/1000$, within a periodic box of size $10\Rc$. The initial velocity field is initialized so that the velocity power spectrum varies with wavenumber as $E_k \propto k^{-2}$. Velocities are scaled to the value set by the turbulent virial parameter $\alpha_{\rm turb} = 2$, which measures the relative importance of turbulence and gravity
\citep{bertoldi_mckee_1992, federrath_klessen_2012}:
\begin{equation}
\alpha_{\rm turb} = \frac{5\sigma_{\rm 1D, cloud}^2\Rc}{G\Mc},
\end{equation}
where $\sigma_{\rm 1D, cloud}$ is the line-of-sight velocity dispersion. The clouds have a uniform magnetic field $B_z$ set by the mass-to-flux ratio $\mu=1.3$:
\begin{equation}
\mu = c_1 \sqrt{\frac{E_{\rm mag}}{E_{\rm grav}}},
\end{equation}
where $E_{\rm mag}$ and $E_{\rm grav}$ are the magnetic and gravitational energy, respectively, and the normalization constant $c_1 \approx 0.4$ \citep{mouschovias_spitzer_1976}. We use a mass resolution of $10^{-3} M_{\odot}$ for gas cells, so each cloud initially consists of $2 \times 10^6$ gas cells. The minimum resolved Jeans length is $\Delta x_J = 36$ au.

Table \ref{table:simulation_initial_conditions} restates the cloud properties in our simulations.  Throughout this study, we present results in terms of the cloud freefall time, defined as 
\begin{equation}
t_{\rm ff} = \sqrt{\frac{3\pi}{32 G \rho_{\rm cloud}}},
\end{equation}
 which is $\tff \approx 2$ Myr. We run the simulations until the clouds are completely dispersed by feedback, which occurs at $\sim 8$ Myr (or $4 \tff$).

\begin{table*} 
\begin{tabular}{ |c c c c c c c| } \hline
 $M_{\rm cloud} (M_{\odot})$ & $R_{\rm cloud}$ (pc) & $\alpha_{\rm turb}$ & $\mu$ & $T_0$ (K) & $\Mc/\Delta m$ & $\Delta x_J$ (au) \\
 \hline
$2 \times 10^3$ & 3 & 2 & 1.3 & 10 & $2 \times 10^6$ & 36 \\
 \hline
\end{tabular}
\caption{Initial conditions of the simulated clouds. Columns are the initial cloud mass, size, virial parameter, mass to magnetic flux ratio, temperature, mass resolution and minimum resolved Jeans length \citep{guszejnov_2022}.}
\label{table:simulation_initial_conditions}
\end{table*}

\subsection{Cosmic Ray Methodology}
\label{subsection:cosmic_rays}

\subsubsection{Cosmic Ray Transport}
\label{subsubsection:cr_transport}

Our study extends the STARFORGE simulation suite by including the {\small GIZMO} module for ``single energy bin'' CRT. This evolves the CR total energy and flux directly, integrating over the CR spectrum, so is effectively representing the $\sim 0.5-10\,$GeV protons, which contain most of the CR energy and pressure. This CR treatment has been used extensively in the FIRE suite of cosmological simulations \citep{chan_2019, hopkins_2020, hopkins_2021a, hopkins_2021b, hopkins_2021c, hopkins_2021d, hopkins_2022a, hopkins_2022b}. The {\small GIZMO} implementation treats CRs as a relativistic fluid ($\gammacr=4/3$) obeying an appropriate set of fluid-like equations derived from the collisionless Vlasov equation. We use the treatment of CRs first formulated in \cite{hopkins_2022a}, which integrates a set of moment equations using a reduced-speed-of-light (RSOL) approximation. This method allows for anisotropic CR transport with streaming, advection, diffusion, and energy losses. We refer readers to \cite{hopkins_2022a} for a detailed derivation of the CRT equations and provide only the necessary details here.

We evolve the total CR energy $\ecr$ and the CR energy flux $\Fecr$; note that $\Fecr$ is a scalar quantity describing the flux parallel to the local magnetic field; the cross-field flux is assumed to be zero. The expressions solved for $\ecr$ and $\Fecr$ are given by
\begin{equation}
D_t \ecr + \nabla \cdot (\Fecr \bhat) =  \Ssc + \Se - \Gammacr - \Gammast -\Pop:\nabla \vgas,
\label{eq:cr_energy_one}
\end{equation}
\begin{equation}
D_t \Fecr + c^2\,\hat{\bf B}\cdot (\nabla \cdot \Pop) = -\barnu \lb \Fecr - 3\chi_e \vst (\ecr+\Pcr) \rb + \Sfe,
\label{eq:cr_energy_two}
\end{equation}
where $\vgas$ is the gas velocity, $\hat{\bf B}$ is the local magnetic field direction, $\Pop = 3 \Pcr \Dop$ is the CR pressure tensor with a scalar pressure $\Pcr \approx \ecr/3$, $\Dop$ is the Eddington tensor, $\barnu = \barnu^+ + \barnu^-$ is the total scattering rate from forward and backward scattering rates in the Alfv\`en frame (related to the CR diffusion coefficient $D_{\parallel} = \vcr^2/3\barnu$ discussed in Section \ref{subsubsection:cr_diffusion_coefficient}), $\vcr\approx c$ is the velocity of an individual CR particle,  and $\chi_e$ is a closure variable (discussed further below). The streaming velocity $\vst$ is approximated as the ion Alfv\'en speed $\vAi=B/\sqrt{4\pi\rho_{\rm ion}}$, where $\rho_{\rm ion} = \chi\rho_{\rm gas} = m_{\rm ion}n f_{\rm ion}$, $\chi=f_{\rm ion}m_{\rm ion}/m_p$ is the mass-weighted ionization fraction, $\rho_{\rm gas}$ is the total gas density, and $m_{\rm ion}$ and $f_{\rm ion}$ are the mean ion mass and ionization fraction by number respectively. \footnote{Formally, the streaming speed is $[(\nu_{+}-\nu_{-})/(\nu_{+}+\nu_{-})]\,\vAi$ where the prefactor represents the fractional asymmetry in the forward/backward scattering rates. But for $\sim$\,GeV in the limit of interest where the streaming speed is large (the case we explore here), this is almost always expected to be $\sim1$.},\footnote{The mean ion mass $m_{\rm ion}$ is approximated as $m_{\rm ion}=m_p$, which is equivalent to setting $\chi = f_{\rm ion}$. This is not accurate in MCs where the dominant charge carrier is $\rm HCO^+$ ($m_{\rm ion}=29 m_p$) \citep{krumholz_2020}; however, since the CRT is calculated assuming $m_{\rm ion}=m_p$, for consistency we make this assumption throughout this study for analysis. We discuss the implications of this in Section \ref{subsubsection:caveats_four}.} The $\Pop:\nabla \vgas$ term in Equation \ref{eq:cr_energy_one} represents the adiabatic compression of CRs with the gas. The term $\Ssc = (\barnu/c^2)3\chi_e \vAi^2 (\ecr+\Pcr)$ represents diffusive reacceleration. The $\Se$ and $\Sfe$ terms represent sources of CR energy and momentum respectively, which in our simulations are only CR injection from stellar winds. The $\Gammacr$ and $\Gammast = -(\barnu/c^2)\vst\Fecr$ terms represent collisional and streaming instability energy losses respectively, which are discussed further in Section \ref{subsubsection:energy_losses}. Finally, the $-\barnu\Fecr$ and $\barnu3\chi_e\vst(\ecr+\Pcr)$ terms in Equation \ref{eq:cr_energy_two} represent scattering  and streaming, respectively, where the latter arises if the scattering is asymmetric. 

The Eddington tensor $\Dop = \chi_e \mathds{I} + (1-3\chi_e) \bhat\bhat$ is defined in terms of the variable $\chi_e = (1-\langle\mu_f^2\rangle)/2$, where a closure relation is used for $\muftwo$ to provide a complete system of equations. \cite{hopkins_2022a} assumes a closure relation analogous to that  adopted in radiation hydrodynamics \citep{levermore_1984}, which is a function that smoothly interpolates between the diffusive and streaming limits:
\begin{equation}
\muftwo \approx \frac{3+4\mufone^2}{5+2(4-3\mufone^2)^{1/2}},
\end{equation}
where $\mufone=\Fecr/\vcr \ecr$. In the diffusive limit, $\mufone=0$ and $\muftwo=1/3$, so $\chi_e=1/3$. In the free streaming limit, $\mufone=1$ and $\muftwo=1$, so $\chi_e=0$.

\subsubsection{Coupled Gas and Cosmic-Ray Equations of Motion}
\label{subsubsection:gas_evolution}

Since CRs exchange energy and momentum with the gas via scattering and Lorentz forces \citep{zweibel_2013, zweibel_2017, hopkins_2022a}, the default {\small GIZMO} momentum and energy equations must be modified to account for the influence of CRs. Spatial differences in the CR pressure tensor $\nabla \cdot \Pop$ produce a net CR current. The current parallel to $\bhat$ is resisted only by exchanging momentum with the gas; therefore, the force balances the scattering + streaming term in Equation \ref{eq:cr_energy_two}. The perpendicular current is redirected by Lorentz forces that exert an equal and opposite force on the gas, which is $\approx -\nabla_{\bot} \cdot \Pop$ \citep{zweibel_2017, hopkins_2022a}where for convenience we have introduced the shorthand notation $\nabla_\parallel \cdot \mathds{Q} \equiv \bhat (\nabla \cdot \mathds{Q})$ and $\nabla_\perp\cdot \mathds{Q} \equiv \nabla\cdot\mathds{Q} - \nabla_\parallel\cdot\mathds{Q}$ for any quantity $\mathds{Q}$. Using Equation \ref{eq:cr_energy_two}, 
the final modified momentum equation can be written as 
\begin{align}
D_t(\rho\vgas) + \ldots+\nabla \cdot\Pop = \nonumber
\\
\bhat(\nabla \cdot\Pop)+ \lp \frac{\bhat}{c^2}\rp\barnu \lb \Fecr - 3\chi_e \vst (\ecr+\Pcr) \rb, 
\label{eq:gas_momentum_equation}
\end{align}
where $\ldots$ refers to all non-CR terms. This has a pressure gradient term and a source term that vanishes when the flux is in steady state.

 The modifications to the energy equation balance the CR energy gains and losses. These include CR energy losses due to work done on the gas outlined in Equation \ref{eq:gas_momentum_equation} along with the four gain and loss terms shown in Equation \ref{eq:cr_energy_one} (energy gained from diffusive reacceleration, CR energy injection, energy lost due to streaming, and collisional and non-collisional losses). Collecting all these terms, the final modified energy equation is 
\begin{equation}
D_t(e_{\rm tot}) + \ldots = \vgas \cdot D_t(\rho \vgas)|_{\rm cr} + \Gammast + \Sgas - \Ssc - \Se ,
\label{eq:gas_energy_equation}
\end{equation}
where $e_{\rm tot} = e_{\rm kin}+e_{\rm mag}+e_{\rm rad}+e_{\rm int}$ is the sum of the gas kinetic, magnetic, radiation, and internal energies, $D_t(\rho \vgas)|_{\rm cr}$ is the CR term on the right side of Equation \ref{eq:gas_momentum_equation}, and $\Sgas$ is the gas heating term due to CR collisional energy losses (described in Section \ref{subsubsection:energy_losses}).

\subsubsection{Energy Loss Processes}
\label{subsubsection:energy_losses}

While they propagate, CRs lose energy due to interactions with non-relativistic particles. Here we adopt an estimate for combined hadronic, Coulomb, and ionization losses. The hadronic/Coulomb losses are from \cite{volk_1996} and \cite{ensslin_1997} and are updated in \cite{guo_oh_2008}. The total CR energy loss rate due to these processes is $\Lambdacr= \Lambda_{\rm coul} + \Lambda_{\rm had} + \Lambda_{\rm ion}$, 
\begin{equation}
\Lambdacr = 7.51 \times 10^{-16}(1+0.22\tilde{n}_e+0.125f_{\rm neut})\lp \frac{n_{\rm H}}{\rm cm^{-3} }\rp \rm s^{-1},
\label{eq:cr_energy_loss}
\end{equation}
where $n_{\rm H}$ is the hydrogen number density, $\tilde{n}_e$ is the number of free electrons per hydrogen nucleus, and $f_{\rm neut}=(1-f_{\rm ion})$ is the neutral gas fraction. 
The energy loss term in Equation \ref{eq:cr_energy_one} is then $\Gammacr=\Lambdacr\ecr$. Following \cite{guo_oh_2008}, we assume 1/6 of the hadronic losses and all non-hadronic losses are thermalized, adding a volumetric heating component to the gas 
\begin{equation}
\Lambda_{\rm gas} = 7.51 \times 10^{-16} \lp0.17 + 0.22\tilde{n}_e + 0.125f_{\rm neut}\rp \lp\frac{n_ {\rm H}}{\rm cm^{-3} }\rp\rm s^{-1}.
\label{eq:lambda_cat_loss}
\end{equation}
The gas heating rate in Equation \ref{eq:gas_energy_equation} is then $\Sgas = \Lambda_{\rm gas}\ecr$.

The CRs also lose energy due to the streaming instability. As they stream, CRs excite Alfv\`en waves caused by anisotropy in the CR distribution. This drives the CRs to stream down the CR pressure gradient and effectively travel at the ion-Alfv\`en speed. The CR energy is transferred to the waves, which are quickly thermalized and heat the gas \citep{kulsrud_pearce_1969, skilling_1975}. CR streaming losses are parameterized in Equation \ref{eq:cr_energy_one} by the term $\Gammast$. We describe the magnitude and impact of these losses in more detail in Section \ref{subsubsection:streaming_instability}.

\subsubsection{Reduced Speed of Light Approximation}
\label{subsubsection:rsol}

Explicitly integrating Equations \ref{eq:cr_energy_one} and \ref{eq:cr_energy_two} imposes a Courant-type time-step limiter $\Delta t \leq C \Delta x/c$, which is numerically prohibitive because $c$ is significantly faster than any other speed in the problem. Therefore, analogously to radiation-hydrodynamics (RHD), we adopt an RSOL approximation, which has been done in many previous CR studies \citep{jiang_oh_2018, chan_2019, hopkins_2020}. Following \cite{hopkins_2022a}, we use an RSOL approximation for Equations \ref{eq:cr_energy_one} and \ref{eq:cr_energy_two} equivalent to replacing $c^{-1}D_t\ecr \rightarrow \tilde{c}^{-1}D_t\ecr$ and  $c^{-1}D_t(\Fecr/c) \rightarrow \tilde{c}^{-1}D_t(\Fecr/c)$ in the original CR transport equation:
\begin{align}
\frac{1}{\tilde{c}}D_t \ecr + \nabla \cdot \lp \frac{\Fecr}{c} \bhat\rp \nonumber
\\
=  \frac{1}{c}\lp\Ssc + \Se -\Gammacr - \Gammast -\Pop:\nabla \vgas\rp,
\label{eq:cr_energy_rsol_one}
\end{align}
\begin{align}
\frac{1}{\tilde{c}} D_t \lp \frac{\Fecr}{c} \rp + (\nabla \cdot \Pop) \nonumber
\\
= \lp \frac{-\barnu}{c^2} \rp \lb \Fecr - 3\chi_e \vst (\ecr+\Pcr) \rb + \lp \frac{1}{c^2} \rp\Sfe.
\label{eq:cr_energy_rsol_two}
\end{align}
We use a value of $\tilde{c} = 300 \rm km/s$ in our simulations, which is justified through a low-resolution convergence study presented in the Appendix.   

The STARFORGE simulations also use an RSOL approximation in the radiation flux equation \citep{grudic_2021}. The default STARFORGE RSOL for radiation is either $30 \rm km/s$ \citep{grudic_2021, grudic_2022} or $90 \rm km/s$ \citep{grudic_2023}; however, in these runs we set $\tilde{c} = 300 \rm km/s$ for radiation for consistency since it does not increase the computational cost. We note that adopting a higher RSOL is primarily important for accurately capturing the optically thick, strong-radiation field limits characteristic of massive star formation \citep{skinner_ostriker_2013}. 

\subsubsection{Cosmic Ray Diffusion Coefficient -  Constant Coefficient Case}
\label{subsubsection:cr_diffusion_coefficient}

The key ingredient in our CR evolution equations that we have not yet specified is the value of the diffusion coefficient $D_\parallel$; the value of this parameter is, as discussed in Section \ref{sec:introduction}, crucial to the outcome. For our fiducial simulations we adopt a diffusion coefficient $D_{\parallel}$ that is constant spatially and temporally. 

We derive the diffusion coefficient following \cite{sampson_2022}, where the authors use an ensemble of MHD turbulence simulations to quantity how the basic parameters describing ISM turbulence affect the transport of streaming CRs. They make the assumption that the total scattering rate $\barnu$ is extremely large (driven by the CR streaming instability) and also asymmetric. In these limits, $\vst \rightarrow \vAi$ (Section \ref{subsubsection:cr_transport}) so that streaming dominates the CR transport and the microphysical diffusion coefficient $D_{\parallel, \rm micro}=c^2/3\barnu$ arising from CR pitch angle scattering is negligibly small in comparison to the effective diffusion coefficient induced by random walk of magnetic field lines (and related processes) due to unresolved turbulence. Under this assumption, they derive a macroscopic diffusion coefficient $D_{\parallel, \rm FLW}$ that quantifies this process; the model is effectively a Large Eddy Simulation closure for CR transport. 

Since $D_{\parallel, \rm FLW}$ does not arise from CR pitch angle scattering, it and $D_{\parallel,\rm micro}$ cannot be formally added together in Equations \ref{eq:cr_energy_one} and \ref{eq:cr_energy_two} (there is no scattering rate $\barnu$ that can be derived from the two that can then be used to obtain a total diffusion coefficient using $\barnu =c^2/3D_{\parallel}$). However, as long as Equation \ref{eq:cr_energy_two} is in steady state ($D_t \Fecr \rightarrow 0$) all terms appear as they should. In our simulations, Equation \ref{eq:cr_energy_two} reaches equilibrium on a timescale $\Delta t_{\rm F} \sim (c/\tilde{c})\barnu^{-1} \sim 10$ yr \citep{hopkins_2021a}, which is much shorter than the macroscopic timescales we consider. Finally, since $D_{\parallel, \rm FLW}$ is derived under the assumption that $D_{\parallel, \rm FLW} \gg D_{\parallel, \rm micro}$ (and we show in Section \ref{subsubsection:variable_cr_diffusion_coefficient} that this is a valid assumption), we arrive at  $D_{\parallel} \sim D_{\parallel, \rm FLW}$.

The diffusion coefficient of interest here is that measured on the minimum size scale $l_\mathrm{min}$ that is resolved by our simulations, since processes on larger scales are presumably resolved directly. Since {\small GIZMO} is a Lagrangian method, the spatial resolution varies throughout the simulation according to the local conditions. Therefore, we estimate $l_{\rm min}$ based on the dynamic range of the cloud turbulent power spectrum.
The minimum resolved length scale for a turbulent sphere is given by $l_{\rm min} = R_{\rm cloud}/k_{\rm max}$, where $k_{\rm max} = N_{\rm 1D}/N_{\rm cells,eff}$ is the maximum wavenumber of the turbulence, $N_{\rm 1D}=(N_{\rm gas}/(4\pi/3))^{1/3}$ is the 1D resolution of the simulation and $N_{\rm cells,eff}$ is the minimum number of cells required to resolve a turbulent mode. For our simulations $N_{\rm gas}= 2 \times 10^6$ and $N_{\rm cells,eff} \approx 10$, which give $l_{\rm min} = 0.38$ pc.

\cite{sampson_2022} present empirical fits for the parallel and perpendicular diffusion coefficient as a function of plasma parameters to be used as sub-grid recipes in simulations. We use the initial cloud magnetic field value $B \approx 7 \mu\rm G$ and average gas density $\rho_{\rm cloud} \approx 18 M_{\odot}/\rm pc^3$. For the ionization fraction, we adopt $\chi = 10^{-7}$, which is within the range of observed values for Milky-Way molecular clouds \citep{caselli_1998, williams_1998}. For the velocity dispersion, we adopt the observed velocity dispersion for turbulent molecular clouds \citep{mckee_2007} 
\begin{equation}
\sigma_{\rm 1D}=0.9\lp\frac{l_{\rm min}}{1 \rm pc}\rp^{0.56} \rm km\,  s^{-1}, 
\end{equation}
which gives $\sigma_{\rm 1D} \approx 0.53$ km/s. Finally, we assume Gaussian diffusion (where the CR distribution function follows $\partial f/\partial t = D_{\rm FLW} \nabla^2 f$), which is predicted for the characteristics of our problem.
Following \cite{sampson_2022}, we obtain values for the parallel and perpendicular diffusion coefficients of $D_{\parallel, \rm FLW} \approx 8.33 \times 10^{25} \rm cm^2/s$ and $D_{\bot, \rm FLW} \approx 4.3 \times 10^{22} \rm cm^2/s$. Since $D_{\bot, \rm FLW} \ll D_{\parallel, \rm FLW}$, we assume $D_{\bot, \rm FLW} = 0$ and adopt $D_{\parallel, \rm FLW}$ as the constant value of the diffusion coefficient.

We note that these derived diffusion coefficients are much lower than the average galactic diffusion coefficient $D_{\rm obs} \sim 10^{28} \rm cm^2/s$ derived from CR data \citep{evoli_2019} for two reasons. First, we resolve CR streaming explicitly in the CR transport equations, while the observed diffusion coefficient arises from both scattering and unresolved streaming ($D_{\rm obs} =D_{\parallel}+ v_{\rm st}\ell_{\rm cr}$). Additionally, $D_{\rm obs}$ is measured in the ionized ISM where $D_{\parallel, \rm FLW} \ll D_{\parallel, \rm micro}$. 
The magnitude of $D_{\parallel, \rm FLW}$ is orders of magnitude smaller when $f_{\rm ion} \sim 1$ compared to where $f_{\rm ion} \sim 10^{-7}$ as in MCs \citep{sampson_2022}; in contrast when $f_{\rm ion} \sim 1$, $D_{\parallel, \rm micro}$ arises from a different pitch-angle scattering mechanism and is much larger. In principle, $D_{\parallel, \rm micro}$ arises from CR pitch angle scattering off both the Alfv\`en waves self-generated by the streaming instability and the extrinsic ISM turbulence ($\barnu = \barnu^{\rm sc}+\barnu^{\rm et}$). \cite{sampson_2022} assume $\barnu^{\rm sc} \gg \barnu^{\rm et}$, and that $\barnu^{\rm sc}$ is so large that the corresponding diffusion coefficient $D_{\parallel, \rm micro} \sim D_{\rm sc}$ is negligibly small; we show in Section \ref{subsubsection:variable_cr_diffusion_coefficient} that this is a reasonable assumption in MCs. In contrast, in higher ionization environments CR scattering is dominated by resonant interactions with extrinsic turbulence \citep{xu_2017, krumholz_2020}, so $\barnu^{\rm et} \gg \barnu^{\rm sc}$ and $D_{\parallel, \rm micro} \sim D_{\rm obs} \sim D_{\rm et}$ \citep{krumholz_2020}. In short, $D_{\rm obs}$ cannot be compared to either the value $D_{\parallel, \rm FLW}$ derived in this section or the value of $D_{\parallel, \rm micro}$ derived in Section \ref{subsubsection:variable_cr_diffusion_coefficient}. 

In addition to our derived value of $D_{\parallel, \rm FLW}$, we explore the impact of varying $D_{\parallel, \rm FLW}$ by an order of magnitude. We also run simulations with a non-uniform CR diffusion coefficient, which we describe in Section \ref{subsubsection:variable_cr_diffusion_coefficient}. Our full suite of simulations, including the diffusion coefficient used for each, is described in Table \ref{table:simulation_cr_initial_conditions}.

\subsubsection{Cosmic Ray Diffusion Coefficient - Variable Coefficient Case}
\label{subsubsection:variable_cr_diffusion_coefficient}

In addition to running simulations with a constant CR diffusion coefficient, we also perform one simulation with a CR diffusion coefficient that varies spatially and temporally. Unlike the diffusion coefficient described in Section \ref{subsubsection:cr_diffusion_coefficient}, which assumes CRT is dominated by streaming and the CR scattering rate $\barnu$ is large ($D_{\parallel, \rm micro} \ll D_{\parallel, \rm FLW}$), this simulation calculates the microphysical diffusion coefficient $D_{\parallel, \rm micro}$ self-consistently based on the local gas and CR properties and neglects field line wandering ($D_{\parallel, \rm FLW} \rightarrow 0$). This has the advantage that it \textit{predicts} whether CRs should be in the streaming dominated limit rather than beginning with that postulate. It should also correctly model `ballistic' CR transport where $\barnu$ is small enough that CR scattering is entirely negligible for the entire CR trajectory and the CRs free stream down field lines through the cloud.  

We use a model for the microphysical diffusion coefficient included in {\small GIZMO} which encapsulates CR pitch angle scattering off both the Alfv\`en waves self-generated by the streaming instability and scattering off extrinsic ISM turbulence ($\barnu = \barnu^{\rm sc}+\barnu^{\rm et}$) \citep{hopkins_2022b}. For the extrinsic turbulence scattering component $\barnu^{\rm et}$, we use the default model from \cite{hopkins_2022b}. We find in our simulations that in low-ionization environments such as starburst galaxies and MCs, efficient ion-neutral damping prevents extrinsic ISM turbulence from cascading down to the scales of CR gyroradii \citep{krumholz_2020}, and so this contribution is largely negligible. 

Instead, CRT is dominated by streaming along field lines at the streaming speed $\vst$, which is the ion-Alf\`ven speed and is modeled directly in the streaming terms in Equations \ref{eq:cr_energy_one} and \ref{eq:cr_energy_two}. In steady state, the wave damping rate $\Gamma_{\rm damp}$ induced by ion-neutral damping and other various damping mechanisms must be matched by the growth rate of the streaming instability $\Gamma_{\rm SI}$ \citep{hopkins_2022b}. The bulk CR propagation speed is $v_D=\vst+v_{\rm diff}=\vAi+v_{\rm diff}$, where $v_{\rm diff}$ is the diffusive velocity component that arises from scattering off the CR-induced Alfv\`en waves and $D_{\rm sc}\propto v_{\rm diff}$ is the corresponding diffusion coefficient. Since $\Gamma_{\rm SI} \propto v_{D}$, an increase in $\Gamma_{\rm damp}$ allows the CRs to propagate faster ($v_{\rm diff}$ and thus $D_{\rm sc}$ increase) until once again $\Gamma_{\rm damp}=\Gamma_{\rm SI}$. In other words, increased wave damping induces a finite scattering rate off the Alfv\`en waves $\barnu^{\rm sc}$ with a diffusion coefficient $D_{\rm sc} = c^2/3\barnu^{\rm sc}$.

In general, the damping rate $\Gamma_{\rm damp}$ includes multiple damping mechanisms including ion-neutral, dust, linear Landau and non-linear Landau damping (with rates given in \citealt{hopkins_2022b}). In practice, here ion-neutral damping dominates and other sources of damping are negligible. The efficiency of ion-neutral damping depends on the local gas and CR properties, and can create orders-of-magnitude variations in the CR diffusivity. The ion-neutral damping rate is $\Gamma_{\rm IN}= \nu_{\rm in}/2$, where $\nu_{\rm in}$ is the collision frequency between ions and neutrals; in a hydrogen-helium plasma this is approximated as 
\begin{equation}
\Gamma_{\rm IN} 
\approx 10^{-9} f_{\rm neutral} \sqrt{\frac{T_{\rm gas}}{1000 \rm K}} \lp \frac{\rho_{\rm gas}} {10^{-24} \rm g/cm^3} \rp \lp \frac{m_p}{m_{\rm ion}} \rp \rm s^{-1},
\label{eq:ion_neutral_damping_rate}
\end{equation}
where $m_p$ is the proton mass, $m_{\rm ion}$ is the mean ion mass, $f_{\rm neutral}= (1-f_{\rm ion})$ is the neutral gas fraction by number, $T_{\rm gas}$ is the gas temperature, and $\rho_{\rm gas} \approx \mu m_p n_{\rm gas}$ is the total gas density with mean molecular weight $\mu$ \citep{kulsrud_pearce_1969, bustard_zweibel_2021, hopkins_2022b}. The growth rate of the streaming instability is \citep{kulsrud_2005}
\begin{equation}
\Gamma_{\rm SI} 
\approx \alpha \Omega_{\rm ion} \lp \frac{n_{\rm CR}}{n_{\rm ion}}\rp
\lp\frac{v_D}{\vAi} -1 \rp,
\label{eq:streaming_instability_growth_rate}
\end{equation}
where $n_{\rm ion} \approx f_{\rm ion}\rho_{\rm gas}/\mu m_p$ is the ion number density, $n_{\rm CR} \approx \epscr/E_{\rm cr}$ is the CR number density for the CR energy density $\epscr$ and characteristic energy of a CR particle $E_{\rm cr}$, $\Omega_{\rm ion}=eB/m_{\rm ion}c$ is the non-relativistic ion frequency, and $\alpha \approx 1$ is a constant that depends on the details of the numerical integrals over non-gyroresonant interactions. Setting $\Gamma_{\rm SI}=\Gamma_{\rm IN}$ and using $f_{\rm ion} \approx n_{\rm ion}/n_{\rm gas}$ and $E_{\rm cr} \sim 1 \rm GeV$,  
\begin{align}
v_{\rm diff} \approx 300 \vAi f_{\rm ion}f_{\rm neut} \sqrt{\frac{T_{\rm gas}}{1000 \rm K}}\lp\frac{\rho_{\rm gas}} {10^{-24} \rm g/cm^3}\rp^2 \nonumber
\\
\times\lp\frac{B}{\mu G}\rp^{-1}\lp \frac{\epscr}{\rm eV cm^{-3}}\rp^{-1}.
\label{eq:v_diff}
\end{align}
The diffusion coefficient $D_{\rm sc}$ is related to the diffusive component of the drift velocity as  $D_{\rm sc} = v_{\rm diff}\ell_{\rm cr}$, where we define the CR gradient scale length $\ell_{\rm cr} =\ecr/\nabla\ecr$. The CR energy gradient $\nabla \ecr$ is computed in {\small GIZMO} using a slope-limited, second order least-squares gradient estimator \citep{hopkins_2015}. Using typical cloud properties in MCs ($f_{\rm ion} \sim 10^{-7}$, $T \sim 10$ K, $\mu \approx 2.33$, $B \sim 10 \mu G$, $n_{\rm gas} \sim 100 \rm cm^{-3}$), $v_{\rm diff} \sim 0.05 \vAi (\epscr/\rm eV cm^{-3})^{-1}$, so $D_{\rm sc} \sim 5\times10^{26}\,(v_{\rm A, ideal}/10\, \rm km\,s^{-1})$$(f_{\rm ion}/10^{-7})^{-1/2}\,(\epscr/\rm eV cm^{-3})^{-1}$ $(\ell_{\rm cr}/\rm pc)\, cm^{2}/s$. The streaming component of the effective diffusion coefficient at larger scales (which we resolve directly in our simulations) is $\vAi\ell_{\rm cr} \sim 10^{28}\,(v_{\rm A, ideal}/10\, \rm km\,s^{-1})$$(f_{\rm ion}/10^{-7})^{-1/2}(\ell_{\rm cr}/\rm pc)\, cm^{2}/s$. Thus, although the streaming instability necessitates CR scattering, the CR transport is still dominated by streaming down field lines unless $\epscr \ll 1 \rm eV/cm^3$.

For our simulations, we set a lower limit of $8.33 \times 10^{24} \rm cm^2/s$ and an upper limit of $10^{34} \rm cm^2/s$ for $D_{\parallel, \rm micro}$. The upper limit is rarely used and makes little difference to our results, but the lower limit prevents regions outside of the cloud from using very low values of $D_{\parallel, \rm micro}$ that would cause the CRs to quickly lose all their energy before propagating into the cloud.Equation \ref{eq:ion_neutral_damping_rate} shows that the ion-neutral damping rate outside the cloud can be low because of the high ionization fraction and low gas density; however, it is likely that in this regime there are other damping mechanisms which we are not accounting for \citep{hopkins_2022b} because of our focus on the denser molecular gas.  

Finally, we note that this model neglects $D_{\parallel, \rm FLW}$; none of our simulations consider both field-line wandering diffusion  ($D_{\parallel, \rm FLW}$) and pitch-angle scattering diffusion ($D_{\parallel, \rm micro}$), since this is not easy to model mathematically. In this section, we have shown that the assumption that streaming dominates the CR transport is valid ($\vst \gg v_{\rm diff}$); however, the assumption that $D_{\parallel, \rm micro} \ll D_{\parallel, \rm FLW}$ used in Section \ref{subsubsection:cr_diffusion_coefficient} is not always the case. For the remainder of this study we refer only to the diffusion coefficient $D_{\parallel}$ rather than distinguishing between the two.  

\subsubsection{Initial CR Configuration}
\label{subsubsection:initial_cr_configuration}

The simulations assume an initially uniform background $\epscr, _{ \rm med}$ for the CR energy density in the diffuse medium outside the cloud. Integrating the CR flux spectrum in the solar neighborhood appropriately for the energy density and ionization rate gives values of $\epscr \approx 1 \rm eV/cm^3$ and $\zeta \approx 1.6\times 10^{-17} \rm s^{-1}$ respectively \citep{cummings_2016, stone_2019}; therefore, {\small GIZMO} assumes a CRIR of $1.6 \times 10^{-17} \rm s^{-1}$ per $\rm eV/cm^3$. We choose our fiducial value of $\epscr, _{ \rm med} = 1 \rm eV/cm^3$ to be consistent with these measurements and previous measurements of CR ionization rates of $\zeta \approx 10^{-17} \rm s^{-1}$ in star-forming regions \citep{dalgarno_2006}.  Inside the cloud, we start with 10 \% of the medium CR energy density ($\epscr, _{\rm cloud} =0.1\rm eV/cm^3$) to reflect probable attenuation in the dense gas \citep{padovani_2009, padovani_2020}. However, low resolution tests show that the simulations progress similarly independently of whether the initial value is $\epscr, _{\rm cloud}=1 \rm eV/cm^3$ or $0.1 \rm eV/cm^3$.

Analytic calculations suggest that the CR energy density should be lower in MCs where the gas density is higher \citep{padovani_2009}; however, analysis of linewidth observations along sightlines towards diffuse clouds have suggested a range of CRIRs that exceed this value \citep{indriolo_2012, indriolo_2015}.  The mean value in the Milky Way is $\zeta \approx 10^{-16} \rm s^{-1}$, an order of magnitude higher than the canonically assumed value \citep{indriolo_2015}. 
 Additionally, higher ionization rates towards more active star-forming regions in the center of our galaxy, other galaxies  \citep{rivilla_2022, behrens_2022, holdship_2022} and towards supernova remnants (SNRs) \citep{ceccarelli_2011} have been observed (see \cite{padovani_2023} for a review). To reflect these uncertainties, we also perform one simulation with initial values of $\epscr, _{ \rm med} =10\rm eV/cm^3$ and $\epscr, _{ \rm cloud} =1\rm eV/cm^3$.

In total, we run six different simulations varying the CR diffusion coefficient and initial CR energy density. These are described in Table \ref{table:simulation_cr_initial_conditions}. We note that run M2e3\_noCRT does not explicitly model CR transport but instead assumes a fixed CR energy density of $\epscr = 1 \rm eV/cm^3$ (or a CRIR of $\zeta \approx 1.7 \times 10^{-17}\,\rm s^{_1}$), which is used only to calculate the heating and cooling of the gas.

\begin{table*}[hbt!]
\begin{tabular}{ |c c c c| } \hline
 Simulation & $M_{\rm cloud} [M_{\odot}]$ & $\epscr,_{\rm med} [\rm eV/cm^3]$ & $D_{\rm ||} [\rm cm^2/s]$ 
 \\
 \hline
M2e3\_E1\_D8e25 & $2 \times 10^3$ & 1 & $8.33 \times 10^{25}$ \\
M2e3\_noCRT & $2 \times 10^3$ & 0 & N/A \\
M2e3\_E10\_D8e25 & $2 \times 10^3$ & 10 & $8.33 \times 10^{25}$ \\
M2e3\_E1\_D8e24 & $2 \times 10^3$ & 1 & $8.33 \times 10^{24}$ \\
M2e3\_E1\_D8e26 & $2 \times 10^3$  & 1 & $8.33 \times 10^{26}$ \\
M2e3\_E1\_DVAR & $2 \times 10^3$ & 1 & Variable \\
 \hline
\end{tabular}
\caption{Details of the variations we use in the initial conditions and physics of our simulations. Columns are the simulation name, cloud mass, initial CR energy density (outside the cloud), and CR diffusion coefficient.}
\label{table:simulation_cr_initial_conditions}
\end{table*}

\section{Results}
\label{section:results}

\subsection{Cloud Evolution}
\label{subsection:cloud_evolution}

\subsubsection{Overview}
\label{subsubsection:overview}

\begin{figure*}[hbt!]
\centering
\includegraphics[width=0.99 \linewidth]{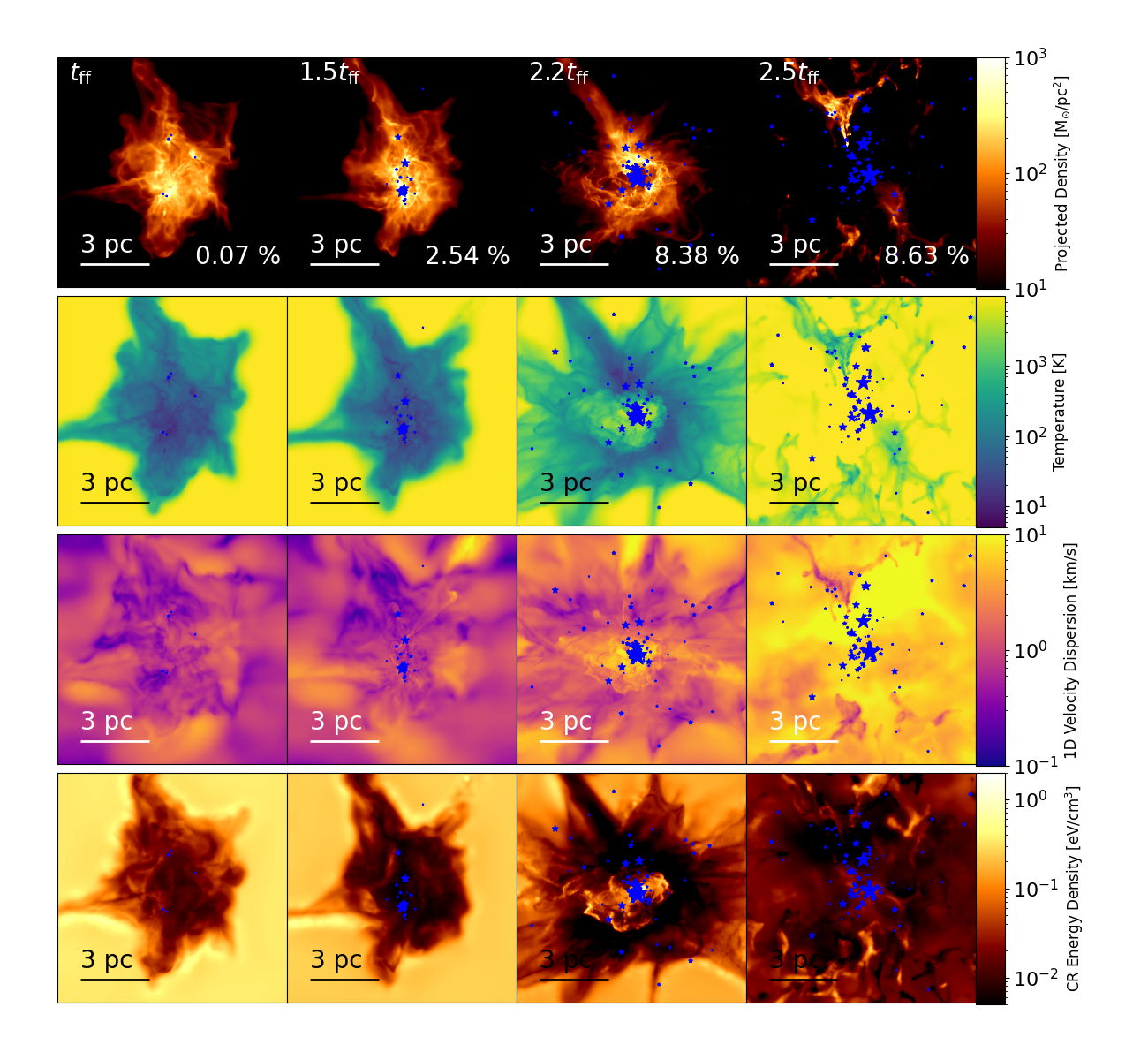}
\caption{Projected density (top row), density-weighted temperature (second row), density-weighted 1D velocity dispersion (third row) and density-weighted CR energy density (bottom row) of the \fiducial\ simulation at four different times between $t_{\rm ff}-2.5t_{\rm ff}$ ($\sim2-5$ Myr). The SFE at each time is given in the bottom right corner. Sink particles are plotted in blue, sized by their mass.}
\label{fig:proj_plots_fid}
\end{figure*}

We begin with a discussion of the cloud evolution for the fiducial cloud including CRT, \fiducial, and a simulation with the same conditions, which assumes a uniform CRIR but no explicit CRT, \nocrs. Figure \ref{fig:proj_plots_fid} shows the time evolution of the projected gas density, density-weighted gas temperature, density-weighted  1D velocity dispersion, and density-weighted CR energy density of the \fiducial\ run, with the positions of stars superimposed in blue. The 1D velocity dispersion for a cell is calculated from the $z$ component of the gas velocity $v_z$ as $\sigma_{1 \rm D}=\sqrt{(v_z-\bar{v_z})^2}$, where $\bar{v_z}$ is the mean value of $v_z$ for all cells along the line of sight. The evolution of the gas in the calculation including CRT follows a similar pattern to that of previous STARFORGE simulations \citep{grudic_2022, guszejnov_2021, guszejnov_2022}. At early times before star formation starts, the initial turbulence in the cloud leads to the formation of dense filaments, which host gravitationally unstable cores. By 1.9 Myr ($\approx \tff$), some of the cores have collapsed and formed the first stars (column 1).

At $\sim 2.9$ Myr ($\approx 1.5 \tff$), the rate of star formation peaks (column 2). The forming protostars launch high-velocity outflows, which slow star formation and stir turbulence in the cloud. By 4.3 Myr ($\approx 2.2 \tff$), massive stars ($> 8 M_{\odot}$) launch stellar winds and produce bubbles of ionizing radiation. This creates a central cavity, which causes the surrounding gas to expand (column 3). By $2.5 \tff$, feedback has completely disrupted the cloud and the cloud is dispersed. 

\begin{figure}[th!]
\centering
\includegraphics[width=0.49 \textwidth]{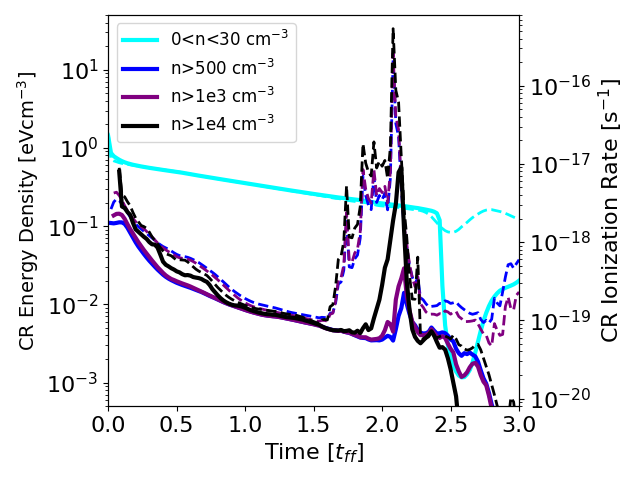}
\caption{Median (solid) and mean (dashed) CR energy density (left axis) and the corresponding CRIR (right axis) for gas in the indicated density range or lower limit for the \fiducial\ simulation. The initial number density inside the cloud is $n_{\rm cloud}\approx 300 \rm cm^{-3}$; therefore, we choose an upper cutoff of 30 $\rm cm^{-3}$ to represent gas outside the cloud. The cutoffs of $500 \rm cm^{-3}$ and $10^{3} \rm cm^{-3}$ are chosen from various definitions to define the cloud boundary, while $10^{4} \rm cm^{-3}$ is a typical density of filamentary substructure and star formation.}
\label{fig:fid_cr_energy_density}
\end{figure}

The bottom row of Figure \ref{fig:proj_plots_fid} follows the evolution of the CR energy density. The cloud boundary is clearly visible because the CR energy losses happen at a faster rate in the cloud material than in the external medium. However, the CR energy density is smoother and exhibits less filamentary substructure than the gas density. The evolution of the CR energy density inside the cloud is shown quantitatively in Figure \ref{fig:fid_cr_energy_density}, which plots the evolution of the median (solid) and mean (dashed) CR energy density as a function of time for various gas densities. Inside the cloud (blue, purple, and black lines), the CR energy density is lower and declines more steeply than outside the cloud (cyan); however, it is largely independent of gas density. The only significant variations before $\sim 1 \tff$ are caused by artificial trapping in the dense gas (Appendix \ref{section:rsol_appendix}), which creates slightly elevated median values in the high density gas (black) \citep{hopkins_2020}. The uniformity of the CR distribution indicates that CRs are losing most of their energy before penetrating into the filaments where stars are forming. 

In all of our simulations massive stars reach the minimum mass for stellar wind launching between $\sim 1.5-2 t_{\rm ff}$. Except for in the \nocrs\ simulation, the massive stars inject CR energy and momentum as they lose mass (Section \ref{subsubsection:starforge_physics}), causing the mean CR energy density in the cloud to rise. Many stars undergo accretion which rapidly increases their mass; consequently, the CR energy injected by each source is nearly constant or increasing.  Figure 2 shows that by $\sim 1.75-2 t_{\rm ff}$ there is enough CR injection to sharply increase the median CR energy density in the cloud, especially in the denser gas where the stars are forming (black). Once the stellar winds rarefy the gas the massive stars are no longer embedded, which causes the CR energy density in the dense gas to decline. 

Finally, the dip in the median CR energy density at low densities (cyan) at $\sim 2.5 \tff$ occurs because the cloud gas disperses faster than the CRs diffuse into this gas. The energy density dips to match the values at $n \gtrsim 500-10^{3} \rm cm^{-3}$ (blue and purple), reflecting the CR levels in the regions previously at these gas densities. Once the background CRs stream into the low density dispersing gas, the median CR energy density rises again.

\begin{figure*}[hbt!]
\centering
\includegraphics[width=0.99 \linewidth]{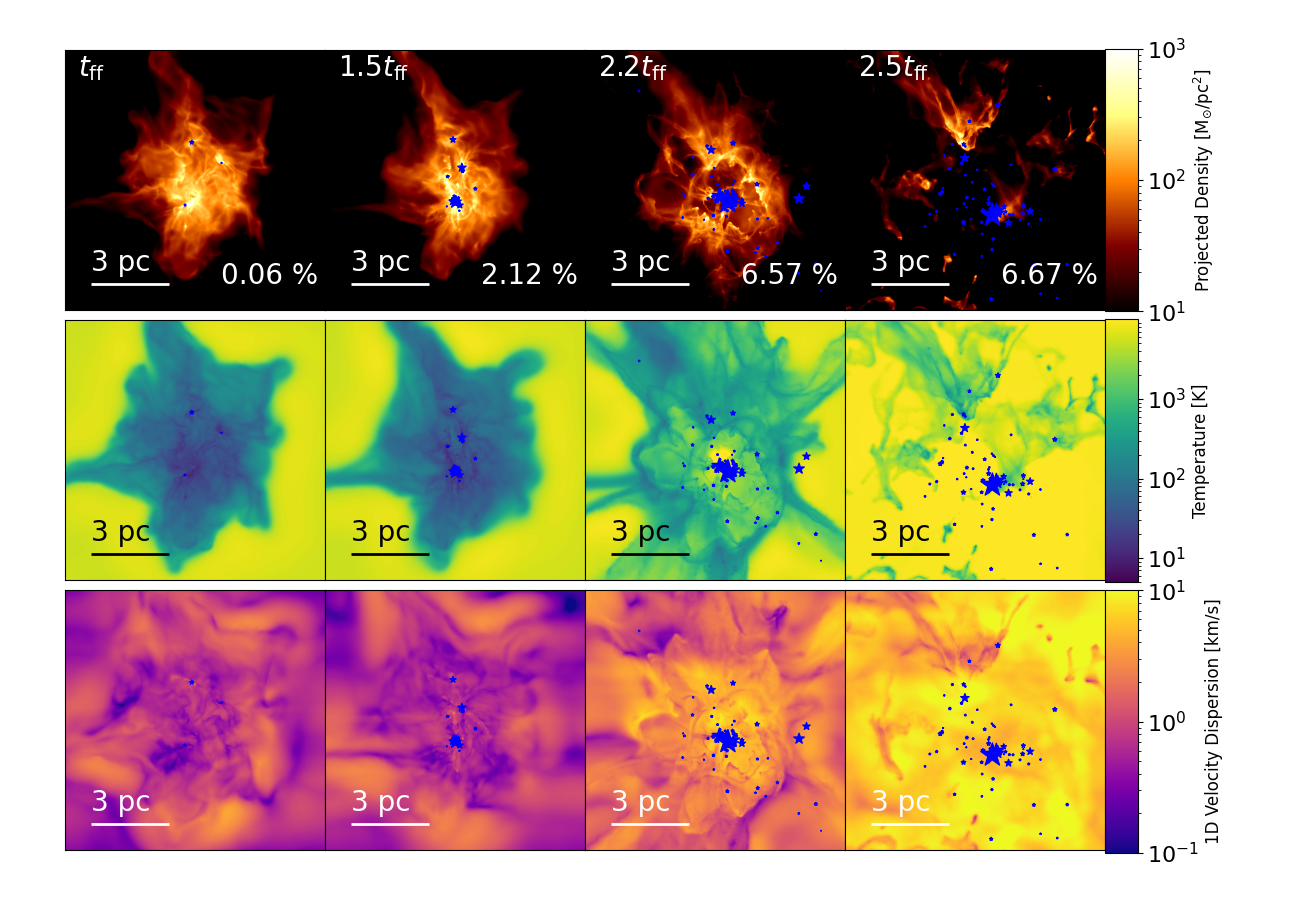}
\caption{Same as Figure \ref{fig:proj_plots_fid} except for the \nocrs\ simulation. In this simulation, a uniform CR energy density of $\epscr= 1 \rm eV/cm^3$ ($\zeta \approx 2 \times 10^{-17} \rm s^{-1}$) is used in the heating and cooling functions.}
\label{fig:proj_plots_noCRs}
\end{figure*}

For comparison, Figure \ref{fig:proj_plots_noCRs} shows the evolution of the gas properties of the \nocrs\ run. In this simulation a uniform CR heating rate is used which 
is equivalent to a CR energy density of $\epscr = 1 \rm eV/cm^3$ ($\Sgas=\Lambda_{\rm gas}\ecr=\Lambda_{\rm gas}\epscr/n_{\rm cr}$ with $\Lambda_{\rm gas}$ given in Equation 8), while in the \fiducial\ run $\epscr < 1 \rm eV/cm^3$ throughout most of the cloud.  Figure \ref{fig:proj_plots_noCRs} shows that despite the differences in the CRIRs between the two runs, the effect of including CRT on the gas structure and overall star formation is minimal. Differences in the gas structure past $\sim$ 4 Myr can largely be attributed to variations in stellar feedback from the different sink particle distributions. 

\begin{figure}[tbh!]
\centering
\includegraphics[width=0.49 \textwidth]{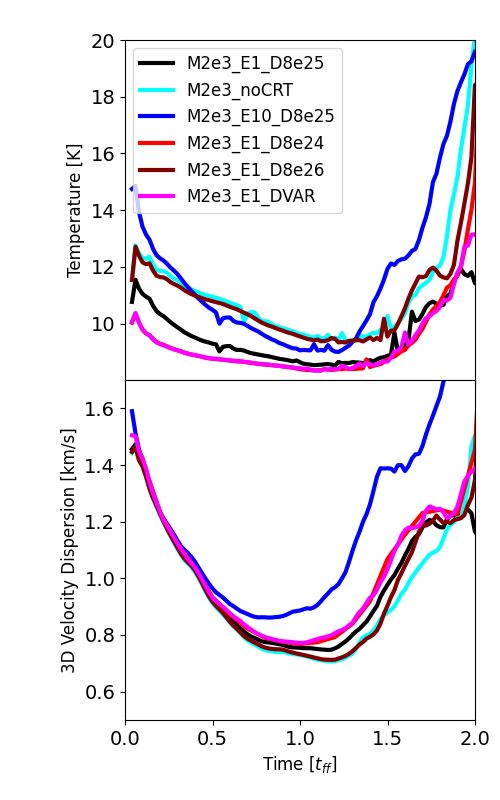}
\caption{Median temperature(top) and median velocity dispersion (bottom) versus time for all gas with number density $n> 1000 \rm cm^{-3}$ for all simulations.}
\label{fig:tff_quantities}
\end{figure}

The effect of including CRT on the gas temperature and velocity dispersion is also minimal. Figure \ref{fig:tff_quantities} shows the evolution of the median temperature (top) and velocity dispersion (bottom) as a function of time at various gas densities for all simulations. The velocity dispersion for a cell is calculated from the three components of the gas velocity $v_x$, $v_y$, and $v_z$ as $\sigma_{3 \rm D}=\sqrt{(v_x-\bar{v_x})^2+(v_y-\bar{v_y})^2+(v_z-\bar{v_z})^2}$, where $\bar{v_x}$, $\bar{v_y}$, and $\bar{v_z}$ are the mean values of $v_x$, $v_y$, and $v_z$ respectively for all cells.
Figure \ref{fig:tff_quantities} shows the  
temperature in the \fiducial\ simulation (black) is actually a couple degrees colder than the \nocrs\ simulation (cyan) due to the lower CR heating.  Similarly, the median velocity dispersion for the two simulations is virtually identical until $t \approx \tff$, indicating CRs are not significantly affecting the gas dynamics in the \fiducial\ run. Beyond $t \approx 1-1.5 \tff$, the cloud begins to disperse due to feedback and differences in these quantities are generally due to minor variations in the stellar mass distribution and cannot be attributed to CRs. 

\subsubsection{Star Formation in a High CR Environment}
\label{subsubsection:high_CRs}

\begin{figure*}[hbt!]
\centering
\includegraphics[width=0.99 \textwidth]{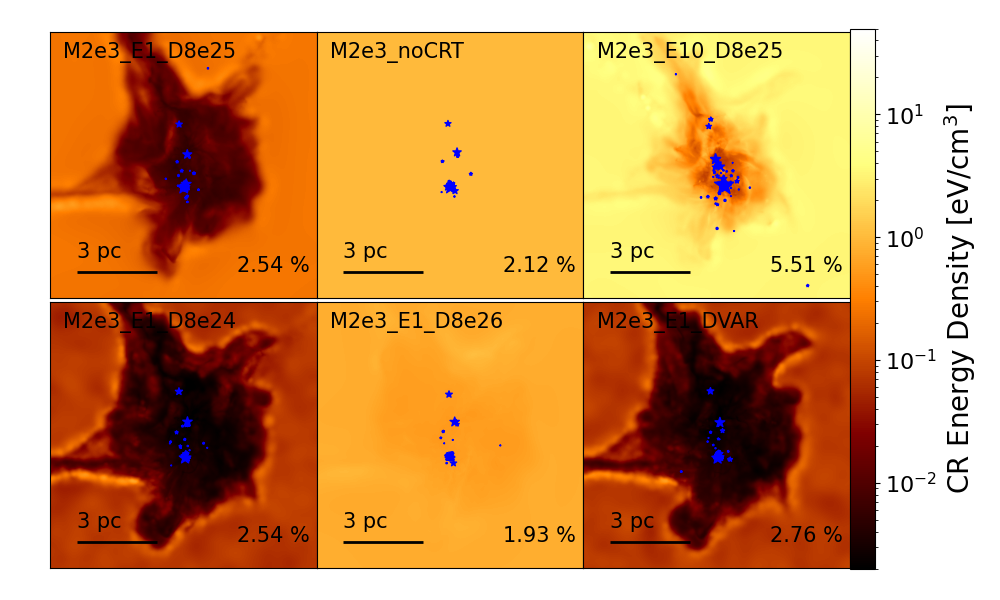}
\includegraphics[width=0.99 \textwidth]{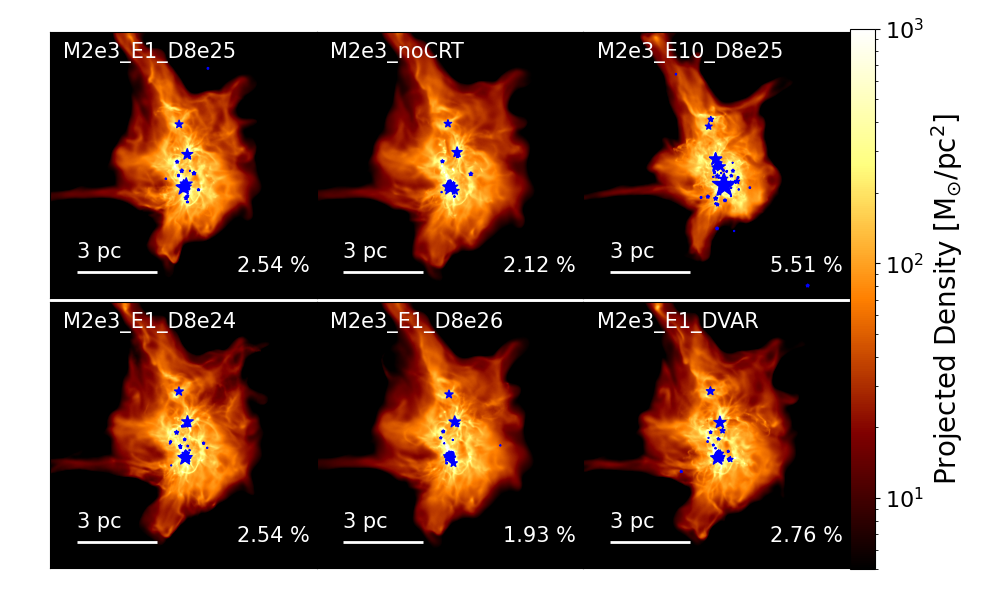}
\caption{Top: Density-weighted CR energy density for all simulations at $t = 1.5 \tff$ ($\approx 3$ Myr). Sink particles are plotted in blue, sized by their mass. The SFE at this time for each simulation is shown in the bottom right corner. For the \nocrs\ panel, a CR energy density of $\epscr=1 \rm eV/cm^3$ is shown, corresponding to the assumed ionization rate of $\zeta \sim 1.7 \times 10^{-17} \rm s^{-1}$. Bottom: Projected gas density for all simulations at $t = 1.5 \tff$ ($\approx 3$ Myr). Sink particle locations are indicated by blue stars, sized by their mass. The current SFE for each simulation is shown in the bottom right corner. Despite significant differences in the CR treatments, the star formation activity and distribution is qualitatively similar.}
\label{fig:simulation_proj_comparison}
\end{figure*}

We now compare the fiducial \fiducial\ CR simulation with the \highcrs\ simulation, which follows a cloud embedded in an environment with a 10 times higher initial CR energy density. The top two rows of Figure \ref{fig:simulation_proj_comparison} show the projected CR energy density for all CRT simulations at $t = 1.5 \tff$ (approximately the time of peak star formation).  The CR energy density in the \highcrs\ cloud declines at a similar rate to that of the \fiducial\ cloud. Despite the higher CR environment, the CR energy density inside the \highcrs\ cloud is relatively low, smaller than the typical 1 eV cm$^{-3}$ generally assumed for clouds. In both runs, by $t = 1.5 \tff$ the  CR energy density in the cloud has declined by over an order of magnitude from its initial value, with values of $\epscr \lesssim 10^{-2} \rm eV/cm^3$ and  $\epscr \lesssim 10^{-1} \rm eV/cm^3$ for the \fiducial\ and \highcrs\ clouds, respectively.

The higher CR energy does, however, affect the rate at which the cloud collapses and the time star formation commences. The bottom two rows of Figure \ref{fig:simulation_proj_comparison} show the projected gas density for all simulations at $t = 1.5 \tff$ ($\approx 3$ Myr). At this time, the \highcrs\ cloud (top right) is more compact and there are more stars compared to the \fiducial\ and \nocrs\ clouds (top left and middle, respectively). This occurs because the relatively high CR energy density outside the cloud acts as an additional pressure on the cloud. While this relative pressure imbalance does not appreciably affect the gas evolution in the \fiducial\ simulation, it is sufficient to accelerate the dynamical evolution of the cloud in the \highcrs\ simulation. Consequently, by $1.5 \tff$ more than twice as much of the cloud mass in the \highcrs\ run has formed (or been accreted onto) stars. 

Although the higher CR background causes a faster initial collapse than in the fiducial \fiducial\ simulation, the other gas properties are qualitatively similar. The top panel of Figure \ref{fig:tff_quantities} shows that the gas in the \highcrs\ simulation cools to a slightly lower median temperature than the \nocrs\ simulation (cyan) once the internal CR energy density declines. However, the cloud temperature is higher after $\sim 1 t_{\rm ff}$ because the elevated star formation causes stronger and earlier heating. The bottom panel shows that the median gas velocity dispersion is also higher after $\sim 0.5 \tff$, a consequence of  the earlier star formation and subsequent impact of feedback within the cloud. Overall, due to the strong CR attenuation in the dense gas, star formation in the high CR environment proceeds relatively similarly to star formation occurring in a typical CR environment. 

\subsubsection{Impact of the CR Diffusion Coefficient on Star Formation}
\label{subsubsection:cr_diffusion_coefficient}

We now turn to a comparison with our simulations with different CR diffusion coefficients. The second row of Figure \ref{fig:simulation_proj_comparison} shows the projected CR energy density at $t = 1.5 \tff$ for the low diffusion coefficient \lowdiffcoeff, high diffusion coefficient \highdiffcoeff, and  variable diffusion coefficient \vardiffcoeff\ simulations. The CR energy density in the cloud is even lower for the \lowdiffcoeff\ (bottom left) run than for the \fiducial\ run, because the CRs scatter more, and consequently suffer greater attenuation by the time they are able to random-walk into the cloud interior. In the \highdiffcoeff\ run (bottom middle), the CRs scatter less frequently and lose comparatively little energy, so the cloud and medium reach an equilibrium value of  $\epscr\approx 1 \rm eV/cm^3$ (second row middle panel of Figure \ref{fig:simulation_proj_comparison}). 

\begin{figure*}
\centering
\includegraphics[width=0.49 \textwidth]{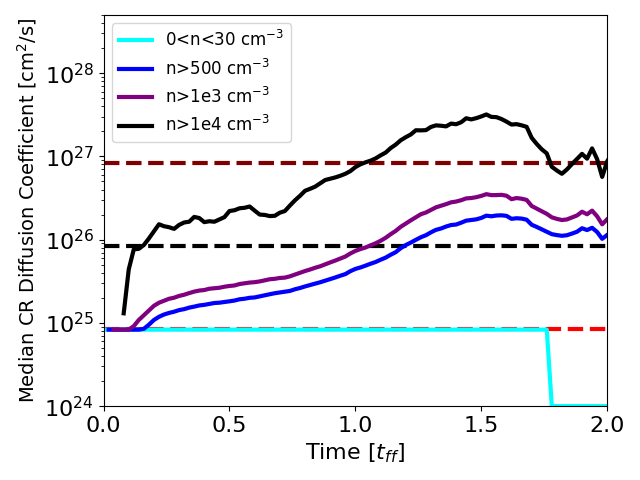}
\includegraphics[width=0.49 \textwidth]{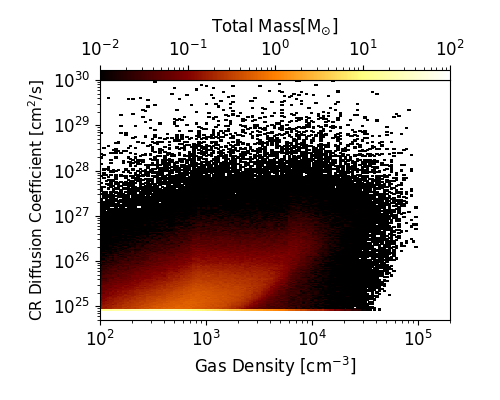}
\caption{Left: Median CR diffusion coefficient 
for gas cells in the indicated density range or lower limit for the \vardiffcoeff\ simulation. Dashed horizontal lines show the constant values of the diffusion coefficient used in the \fiducial\ (black), \lowdiffcoeff\ (red), and \highdiffcoeff\ (brown) simulations. The drop in the median value at $\approx 1.75 \tff$ is due to a numerical error in setting the lower cutoff; however, since this occurs after the cloud has collapsed and the CRs have lost most of their energy it does not affect our results. Right: Total gas mass as a function of gas density and CR diffusion coefficient $D_{\rm sc}$ (not including resolved streaming, Section \ref{subsubsection:variable_cr_diffusion_coefficient}) for the \vardiffcoeff\ simulation at $t=0.5 t_{\rm ff}$.}
\label{fig:var_cr_diff_coeff}
\end{figure*}

The right panel in the second row of Figure \ref{fig:simulation_proj_comparison} shows that the CR energy density in the  \vardiffcoeff\ simulation evolves qualitatively similarly to the \lowdiffcoeff\ simulation. Computing the microphysical diffusion coefficient $D_{\parallel, \rm micro}$ typically gives lower average values than the value of $D_{\parallel, \rm FLW}$ used in the \fiducial\ simulation. Figure \ref{fig:var_cr_diff_coeff} (left) shows the median diffusion coefficient at different gas densities in the \vardiffcoeff\ simulation. 
Outside the cloud (cyan line), the low gas densities and high ionization result in inefficient ion-neutral damping
(Equation \ref{eq:ion_neutral_damping_rate}) so the diffusion coefficient is typically equal to the minimum allowed value. Inside the cloud (purple and blue lines), the values of $D_{\parallel}$ increase as the cloud collapses because the higher densities create more efficient ion-neutral damping, which damps the Alfv\`en waves and decreases the CR scattering rate. The right panel of Figure \ref{fig:var_cr_diff_coeff} shows a 2D histogram of the amplitude of the CR diffusion coefficient as a function of gas density at $t=0.5 t_{\rm ff}$. Before $\sim 1 \tff$ the diffusion coefficient throughout most of the cloud is between the \fiducial\ value and the \lowdiffcoeff\ value except at the highest gas densities. At densities $n<1000 \rm cm^{-3}$, a large fraction of the simulation space computes  our lower cutoff value $D_{\parallel} = 8.33 \times 10^{24} \rm cm^2/s$, which we impose due to the additional contribution of multiple wave damping mechanisms at lower gas densities (Section \ref{subsubsection:variable_cr_diffusion_coefficient}). For most of the cloud collapse, the high scattering rate causes the CRs to lose energy as in the \lowdiffcoeff\ simulation.   

For all three of these variations, the effect of CRT on the cloud evolution and gas properties is minimal. The bottom row of Figure \ref{fig:simulation_proj_comparison} shows that the gas density at $t=1.5 \tff$ looks virtually identical to the gas density of the \fiducial cloud. As expected, the temperature of the \lowdiffcoeff\ and \vardiffcoeff\ simulations shown in Figure \ref{fig:tff_quantities} (red and magenta, respectively) is lower than for the \fiducial\ simulation. The temperature in the \highdiffcoeff\ simulation is similar to that of the \fiducial\ simulation, because both have a value of $\epscr \approx 1 \rm eV/cm^3$ throughout the cloud. The velocity dispersion for all three is comparable, since star formation proceeds similarly. 

\subsection{Star Formation Properties}
\label{subsection:star_formation}

\subsubsection{Star Formation History of the Cloud}
\label{subsubsection:sf_history}

\begin{figure*}
\centering
\includegraphics[width=0.49 \textwidth]{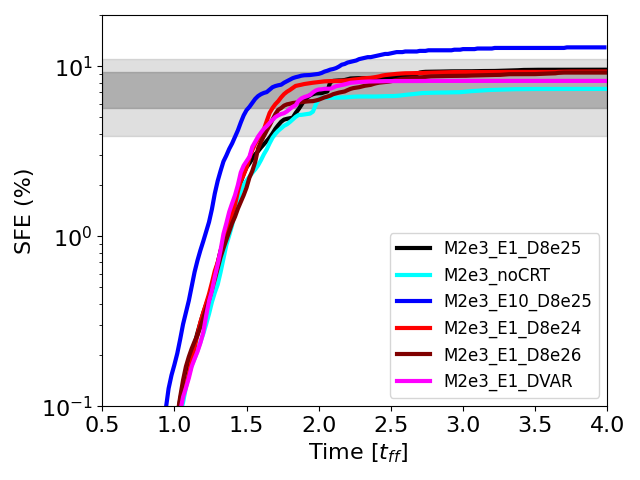}
\includegraphics[width=0.49 \textwidth]{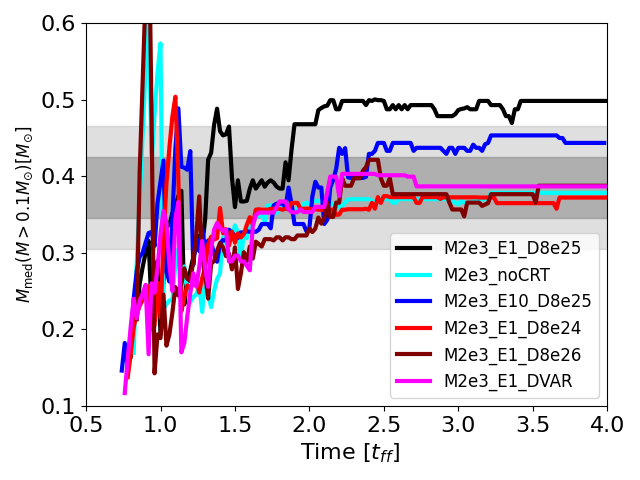}
\caption{Left: Evolution of the star formation efficiency (SFE) versus time for all simulations. Shaded regions are the $1\sigma$ (dark gray) and $2\sigma$ (light gray) final SFE values obtained from a set of 100 2,000 M$_\odot$ cloud simulations without CRT that vary the initial turbulent seed. Right: Median mass for $M_* \ge0.1 M_{\odot}$ versus time for all runs. Shaded regions are the $1\sigma$ (dark gray) and $2\sigma$ (light gray) final $M_{\rm med}$ values obtained from a set of previous simulations run without CRT, varying the initial turbulent seed.}
\label{fig:sfe_mmed}
\end{figure*}

We now turn to various star formation metrics that can be obtained from the simulations and compared to observations. The star formation efficiency (SFE) is defined as the percentage of the initial cloud mass that has been converted to stars:
\begin{equation}
{\rm SFE} (t) = M_*(t)/M_0,
\end{equation}
where $M_* (t)$ is the total stellar mass at time $t$. 
Figure \ref{fig:sfe_mmed}, left, shows the evolution of the SFE for all runs. Qualitatively, the SFE evolves similarly in all cases. Once star formation starts, the SFE undergoes a period of rapid growth as the cloud collapses and the densities increase. The rate of star formation peaks at $\approx 1.5 t_{\rm ff}$. At this time, feedback begins to disperse the cloud, causing the slope of the SFE to decrease. By $\approx 2 \tff$, the SFE has plateaued and the cloud is dispersing. 
 
Figure \ref{fig:sfe_mmed} shows that the inclusion of CRT does not have a significant impact on the star formation history of the cloud for any of our simulations with $\epscr = 1 \rm eV/cm^3$. Up until $\approx 1.5 \tff$, the SFE looks very similar for all of these runs, and minor variations after this can be attributed to differences in star formation feedback. The SFE in the \highcrs\ cloud (blue lines) begins the period of steep rise slightly earlier than the other clouds due to the external CR pressure on the cloud discussed in Section \ref{subsubsection:high_CRs}; however, once star formation starts it proceeds at roughly the same rate for all of the clouds once the background CR level has declined. For all of our clouds, variations in the early evolution of the SFE are smaller than variations caused by different initial turbulent realizations with the same physics \citep{guszejnov_2022}. 

To evaluate the significance of the variation between runs, we compare our results to a suite of 103 simulations of $M=2,000$ M$_\odot$ clouds evolved with different turbulent seeds and no CRT. 100 of these clouds were run with $E_{\rm mag}/E_{\rm grav} = 0.01$ and were presented in \cite{grudic_2023}, while the other three were run with $E_{\rm mag}/E_{\rm grav} = 0.1$ but are otherwise identical. We take the mean value of the final SFE from the three higher-magnetization clouds plus our \nocrs\ run as the `standard' final SFE with no CRT, and assume the standard deviation in the final SFE calculated from the suite of 100 lower magnetization clouds is representative of the variation. The $\pm 1\sigma$ region ($\rm SFE  \approx 7.42 \pm 1.77$), is indicated as a grey shaded area in the left panel of Figure \ref{fig:sfe_mmed}. 

The second column of Table \ref{table:final_sf_parameters} shows the  SFE at the final time $4 \tff$. The final SFE of our \nocrs\ cloud is the lowest. All of our $\epscr = 1 \rm eV/cm^3$ clouds have slightly higher values, which could be due to the extra confining CR pressure; the \highdiffcoeff\ and \vardiffcoeff\ are within 1$\sigma$ while the \fiducial\ and \lowdiffcoeff\ clouds are within 2$\sigma$. The \highcrs\ cloud has a more than 2$\sigma$ greater final SFE. 

\subsubsection{Final Stellar Mass Distribution}
\label{subsubsection:imf}

Figure \ref{fig:sfe_mmed}, right, shows the evolution of the median mass of all sink particles with $M> 0.1 M_{\odot}$. As in the left panel, the shaded region shows the mean of the final values from the four $E_{\rm mag}/E_{\rm grav}=0.1$ simulations (including the \nocrs\ run) with the $1\sigma$ region from the 100 $E_{\rm mag}/E_{\rm grav}=0.01$ simulations. There is no significant difference in the evolution of $M_{\rm med}$ between the CRT and \nocrs\ simulations, except for the \fiducial\ simulation. The median mass for that run experiences a sharp jump between $\sim 1.25-1.5 \tff$ and then rises again sharply at $\sim 1.8 \tff$. While the earlier jump may be due to statistical variation, the latter may be correlated with the CR injection by stellar winds discussed in Section \ref{subsection:cloud_evolution}, which was strongest in our \fiducial\ simulation. It occurs at approximately the same time as the sharp increase in the median CR energy density in the cloud from stellar wind injection shown in Figure \ref{fig:fid_cr_energy_density}. The additional CR energy may affect the gas temperature and pressure, which would effect the stellar mass distribution. A larger statistical study is required to draw firm conclusions. 

The final value of $M_{\rm med}$ at $t=4 \tff$ is shown in the third column of Table \ref{table:final_sf_parameters}. Except for the \fiducial\ simulation, all of our $\epscr=1 \rm eV/cm^3$ runs are within $1\sigma$. The \highcrs\ simulation is slightly higher, within $2\sigma$. 

\begin{figure*}
\centering
\includegraphics[width=0.49 \textwidth]{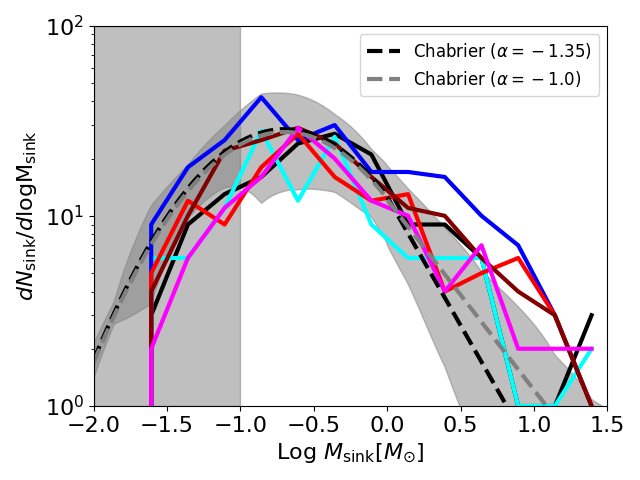}
\includegraphics[width=0.49 \textwidth]{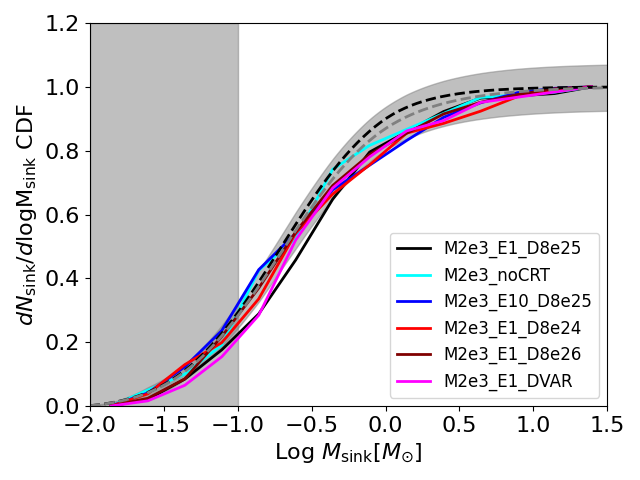}
\caption{Left: Stellar initial mass function (IMF) for all runs at $t = 4\tff$. The shaded region at $M<0.1 M_{\odot}$ indicates the low-mass incompleteness region \citep{grudic_2021}. Dashed lines show the Chabrier IMF \citep{chabrier_2005} with the standard slope of $\alpha=-1.35$ (black) and the slope obtained from previous STARFORGE simulations, $\alpha=-1$ (gray) \citep{grudic_2022, guszejnov_2022}. Shaded region shows the $1 \sigma$ values in each mass bin obtained from a set of 100 2000 $M_{\odot}$ cloud simulations without CRT. Right: Normalized CDF of the IMF for all simulations.}
\label{fig:imf}
\end{figure*}

At the final time, the $\epscr=1 \rm eV/cm^3$ simulations have each formed $\sim 120-170$ stars. Figure \ref{fig:imf} shows the stellar mass distributions at $4 \tff$ (left) and their normalized cumulative distribution functions (right). The shaded region represents the $1 \sigma$ region in each mass bin from the 100 $E_{\rm mag}/E_{\rm grav}=0.01$ simulations for a mass distribution with 136 stars and a high mass slope $\alpha=-1$ in $dN/d\ln M$, shown by the gray dashed line. Qualitatively, the IMF for all of the $\epscr=1 \rm eV/cm^3$  simulations appears reasonably well described by a Chabrier IMF \citep{chabrier_2005}. Prior work indicates that protostellar jets are responsible for setting the mass scale of the IMF \citep{guszejnov_2021}. The high mass slope is flatter than the canonical Chabrier value of $\alpha=-1.35$ but is consistent with previous STARFORGE simulations, which find $\alpha=-1$, possibly due to some missing physics, such as disk mediated accretion and fragmentation, that enhance the number of massive stars \citep{grudic_2022, guszejnov_2022}. The peak of the \fiducial\ simulation is higher than that of the other simulations because of the higher median mass (Figure \ref{fig:sfe_mmed}). Except for this, we find no qualitative difference between the IMFs of the $\epscr=1 \rm eV/cm^3$ CRT simulations and the IMF of the \nocrs\ simulation. 

The \highcrs\ simulation produced twice as many stars as the other simulations ($\sim$ 220). Figure \ref{fig:imf} shows that the number of stars formed with $M \lesssim 0.1 M_{\odot}$ is larger than the other simulations, resulting in a visual peak in the IMF at $\sim 0.1 M_{\odot}$ rather than $\sim 0.3 M_{\odot}$.  
However, the median mass of the \highcrs\ simulation above $0.1 M_{\odot}$ is higher than all of the other simulations except the \fiducial\ simulation (Figure \ref{fig:sfe_mmed}) because this simulation also formed more high-mass stars. Indeed, the high-mass slope of the \highcrs\ simulation in Figure \ref{fig:imf} appears flatter than $\alpha=-1$ and the distribution is well outside the $1 \sigma$ region at $M > 1 M_{\odot}$. Higher resolution simulations with a larger statistical sample of stars are required to confirm this trend.

\begin{table*} 
\begin{tabular}{ |c c c c| } \hline
 Simulation Name & SFE (\%) & $M_{\rm med} (M > 0.1 M_{\odot})$ & $M_{\rm max}$
 \\
 \hline
M2e3\_E1\_D8e25 & 9.53 & 0.50 & 29.3 \\
M2e3\_noCRT & 7.35 & 0.38 & 30.6  \\
M2e3\_E10\_D8e25 & 12.90 & 0.44 & 22.5  \\
M2e3\_E1\_D8e24 & 9.30 & 0.37 & 23.3  \\
M2e3\_E1\_D8e26 & 9.17  & 0.38 & 22.7 \\
M2e3\_E1\_DVAR & 8.19 & 0.39 & 19.1  \\
 \hline
\end{tabular}
\caption{Final values for the SFE, median mass above the incompleteness region, and maximum mass for all simulations at $4 \tff$.}
\label{table:final_sf_parameters}
\end{table*}

\subsection{Ionization}
\label{subsection:ionization}

\begin{figure*}
\centering
\includegraphics[width=0.99 \textwidth]{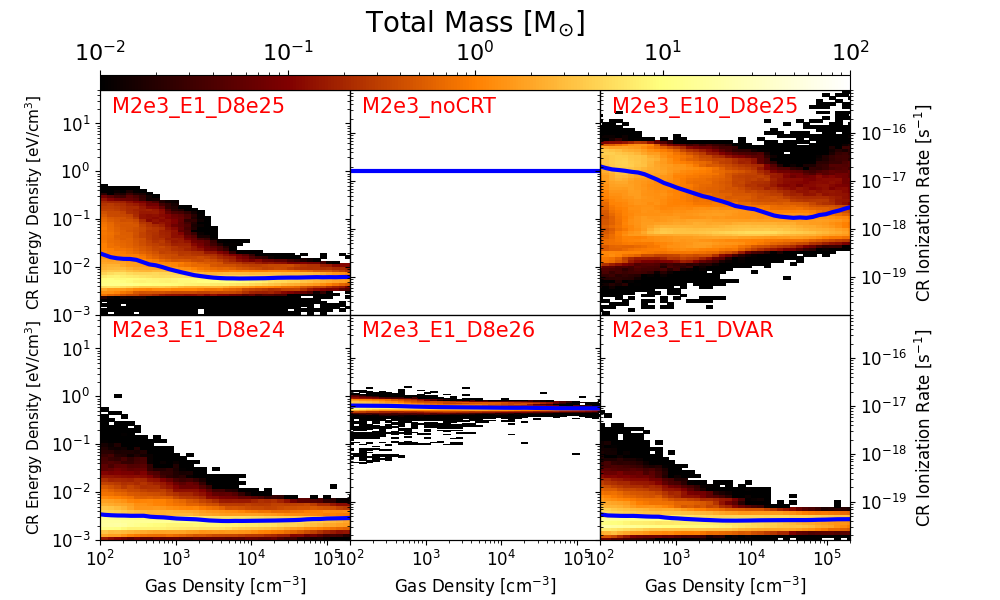}
\caption{The total mass of cells as a function of gas density and CR energy density for all simulations at $t =1.5 \tff$ ($\approx 3$ Myr). The blue line shows the mass-weighted mean CR energy density at each gas density. On the right hand is the corresponding CR ionization rate. The horizontal blue line in the \nocrs\ panel is the MW CR energy density ($\epscr = 1 \rm eV/cm^3$) assumed when CRT is not used.}
\label{fig:cr_energy_density_phase}
\end{figure*}

As shown in Figure \ref{fig:simulation_proj_comparison}, we find that in a Milky Way like environment, for low to moderate values of the CR diffusion coefficient the CRIR is roughly uniform throughout the cloud. The median value ranges between $\zeta \approx 4 \times 10^{-20}  - 2 \times 10^{-19} \rm s^{-1}$ during the main star forming epoch of the simulations ($t \approx 1-1.5 \tff$).  Figure \ref{fig:cr_energy_density_phase} shows a phase plot of the total simulation mass as a function of the CR energy density and gas density at $1.5 \tff$. The CRIR is shown on the right, which we calculate from the CR energy density assuming an ionization rate of $1.6 \times 10^{-17} \rm s^{-1}$ per $\rm eV/cm^3$ \citep{cummings_2016}. The blue line shows the mass-weighted mean CR energy density and corresponding CRIR at each gas density. The \fiducial, \lowdiffcoeff\, and \vardiffcoeff\ clouds have a CRIR that strongly peaks at a value of $\sim 0.5-1 \times 10^{-19}$s$^{-1}$, where some regions with $n< 10^4$ reach $\zeta \approx 10^{-18}$s$^{-1}$. The \highdiffcoeff\ run  has a nearly uniform CRIR of $\zeta \sim 10^{-17}$s$^{-1}$ throughout the cloud. 

The \highcrs\ simulation has elevated ionization rates throughout the cloud compared to the \fiducial\ simulation. The median value of the CRIR ranges from $\approx 1-6 \times 10^{-18} \rm s^{-1}$ between 1-1.5 $\tff$. Although the average CR energy density declines over time in both the \highcrs\ and \fiducial\ clouds, the CR distribution is more variable throughout the \highcrs\ cloud. Figure \ref{fig:cr_energy_density_phase} shows that the typical cloud gas in the \fiducial\ and \highcrs\ simulations has a characteristic ionization rate of $\zeta \approx 10^{-19} \rm s^{-1}$ and $2 \times 10^{-18} \rm s^{-1}$, respectively, relatively independent of gas density. The blue line indicates that the \fiducial\ cloud is strongly peaked at this value, especially at $n \gtrsim 10^4 \rm cm^{-3}$. In contrast, the \highcrs\ cloud has an average CRIR between $\approx 10^{-18} - 10^{-17} \rm s^{-1}$ depending on gas density, indicating that the distribution is not strongly peaked and has a range of values above and below the median CRIR. There are regions with $\zeta \gtrsim 10^{-17} \rm s^{-1}$ at all gas densities. 

\begin{figure*}
\centering
\includegraphics[width=0.99 \textwidth]{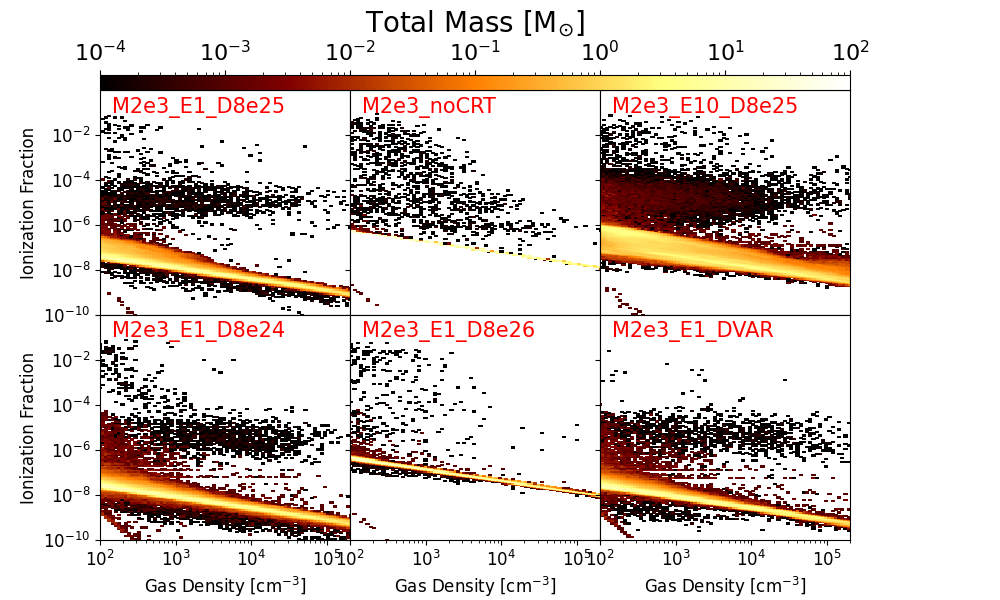}
\caption{Total mass of cells as a function of gas density and ionization fraction for all simulations at $t =1.5 \tff$ ($\approx 3$ Myr). The yellow regions in each panel correspond to CR ionization, while the dark patches are from radiative feedback. The pale yellow line in the \nocrs\ panel shows the contribution assuming a fixed CRIR of $1.6 \times 10^{-17} \rm eV/cm^3$, which declines with gas density as $1/\sqrt{n_H}$. The descending line in all CRT simulations with ionization fraction $<1e-8$ and density $< 1e3$ is due to a known bug in GIZMO where one particle per MPI rank is assigned $\epscr = 0$. Given the fact that we use $2 \times 10^6$ cells and only $\sim 200$ MPI ranks, this bug impacts very few cells and thus does not have an effect on our results.}
\label{fig:ionization_fraction_phase}
\end{figure*}

Figure \ref{fig:ionization_fraction_phase} shows a phase plot of the ionization fraction for all runs at $t=1.5 \tff$. We compute the ionization fraction following the convention used in {\small GIZMO} from the electron fraction $f_e$, neutral hydrogen fraction $f_{\rm H}$, and HII fraction $f_{\rm HII}$ as MAX($f_e, (1-f_{\rm H})$, $f_{\rm HII}$). At this time, ionization produced by stellar feedback is still minimal, so the degree of ionization reflects the local CR energy density. The pale yellow line in the \nocrs\ panel shows the contribution assuming a fixed CRIR of $1.6 \times 10^{-17} \rm s^{-1}$,  which is used in the \nocrs\ simulation. The dark patches in the upper left are produced by ionization caused by radiative feedback, while the area in the lower left shows the contribution from CR ionization. Since the CRIR is low for most of the runs, the contribution from CRs to the overall gas ionization is minor compared to the ionization caused by radiative feedback once star formation is underway.

\section{Discussion}
\label{section:discussion}

\subsection{Propagation of CRs in Molecular Clouds}
\label{subsection:cr_prop_disc}

In this section we discuss the implications of our results for the propagation of CRs within MCs.

\subsubsection{Role of the CR-Driven Streaming Instability}
\label{subsubsection:streaming_instability}

We find that for all of the simulations except the \highdiffcoeff\ simulation, the CR energy density in the cloud is significantly lower than it is outside the cloud. The CR energy density in the cloud declines quickly as the cloud collapses and is not replenished by CRs from outside the cloud. To understand this result, we consider the relevant timescales for CR propagation and compare them to the loss timescales. The following expressions describe the timescale for each transport mechanism to propagate CRs a distance $L$ \citep[e.g.,][]{habegger_2023}:
\begin{equation}
\tau_{\rm adv} = L/v_{\rm adv} \approx 1 L_{\rm pc} v_{\rm km}^{-1} \rm Myr,
\label{eq:tau_adv}
\end{equation}
\begin{equation}
\tau_{\rm str} = L/\vst \approx 1 L_{\rm pc} B_{\mu}^{-1}\rho_{-21}^{1/2}f_{-7}^{1/2} \rm kyr, 
\label{eq:tau_str}
\end{equation}
\begin{equation}
\tau_{\rm diff} = L^2/D_{||} \approx 3 L_{\rm pc}^2 D_{26}^{-1} \rm kyr,
\label{eq:tau_diff}
\end{equation}
for advection, streaming, and diffusion, respectively, where we assume $v_{\rm adv}=|\bold{u}|$ (the gas velocity), $v_{\rm st}=\vAi$, and $L_{\rm pc}, v_{\rm km}, B_{\mu}, \rho_{-21}, f_{-7}$ and $D_{26}$ are the length scale, advection velocity, root-mean-square magnetic field, gas density, ionization fraction, and the diffusion coefficient in units of 1 pc, 1 km/s, $1 \mu G$, $10^{-21} \rm g/cm^3$, $10^{-7}$, and $10^{26} \rm cm^2/s$, respectively.  

While they propagate, the CRs lose energy through streaming and collisional losses. If the amount of time for CRs to lose their energy through either one of these mechanisms is considerably shorter than the time for the CRs to propagate through the cloud, then the CRs will quickly lose energy, and thus the CR energy density will be low inside the cloud. The energy loss timescale from collisions is obtained from Equation \ref{eq:cr_energy_loss}: 
\begin{equation}
\tau_{\rm cat, loss} = \Lambdacr^{-1} =  \lp 7.51 \times 10^{-16} \rp ^{-1 }\lp\frac{n_H}{\rm cm^{-3}}\rp^{-1} \rm s.  
\label{eq:tau_cat_loss}
\end{equation}
The streaming loss term is calculated consistently in the simulations as $\Gammast= -(\barnu/c^2)\vst\Fecr$, where $\Gammast=\Lambda_{\rm st}\ecr$ and $\Lambda_{\rm st}$ is the streaming energy loss rate. The streaming loss timescale can be approximated as  \citep{chan_2019}:
\begin{equation}
\tau_{\rm str, loss} =  \Lambda_{\rm st}^{-1}=\lp\frac{1}{\gammacr-1} \rp | \hat{B} \cdot \hat{\nabla}\epscr|^{-2} \lp\frac{\ell_{\rm cr}}{\vst} \rp \rm s,   \label{eq:tau_str_loss}
\end{equation}
where $\hat{\nabla}\epscr$ is the unit vector of the CR energy density gradient $\nabla\epscr$ and $\ell_{\rm cr}=\epscr/\nabla\epscr$ is the CR gradient scale length.

The advection timescale (Equation \ref{eq:tau_adv}) reflects the speed of the non-relativistic gas and thus represents the gas crossing time in the simulation. Since our simulations use a comoving Lagrangian frame which moves with the gas, it is independent of the choice of $\tilde{c}$. The four other timescales ($\tau_{\rm str}$, $\tau_{\rm diff}$, $\tau_{\rm cat, loss}$ and $\tau_{\rm str, loss}$), however, represent CR propagation and losses away from the gas and thus these timescales are a factor $\Gamma_{\rm RSOL}=c/{\tilde{c}}$ slower due to the RSOL approximation \citep{hopkins_2022a}. For our value of $\tilde{c} = 300$ km/s, $\Gamma_{\rm RSOL}=1000$. This caveat does not influence the conclusions of this section because we are concerned with the relative importance of these timescales and all of the timescales scale proportionally. We discuss the implications of the choice of $\tilde{c}$ further in Section \ref{subsection:simulation_caveats} and in Appendix \ref{section:rsol_appendix}. 

Figure \ref{fig:timescale_quantities} (bottom) shows the five timescales as a function of gas density for the \fiducial\ (solid), and \highdiffcoeff\ (dashed) simulations at $0.5 \tff$, where the relevant gas quantities used to calculate the transport and loss timescales are displayed in the top panel. To reflect the behavior of the simulations we scale all timescales except for $\tau_{\rm adv}$ by a factor of $\Gamma_{\rm RSOL}$. For the \fiducial\ simulation transport due to streaming dominates over advective and diffusive transport throughout the cloud. This is especially true at the highest densities ($n \gtrsim 100 \rm cm^{-3}$), where the ionization fraction is $f_{\rm ion}< 10^{-7}$, resulting in $\vAi > 1000$ km/s. The high streaming velocity also produces efficient energy losses from the streaming instability. Streaming losses dominate for gas densities $n< 10^4 \rm cm^{-3}$ and collisional energy losses dominate at higher densities. Although the magnitudes of the transport and loss timescales vary with the CR and gas properties, all simulations except the \highdiffcoeff\ simulation have significant streaming instability energy losses, which cause the CRs to  lose a significant amount of energy before propagating far in the cloud. To provide additional evidence for this result, we run low-resolution tests where we disable collisional losses and streaming transport and losses separately (see Appendix \ref{section:nost_appendix}).

\begin{figure}[th!]
\centering
\includegraphics[width=0.49 \textwidth]{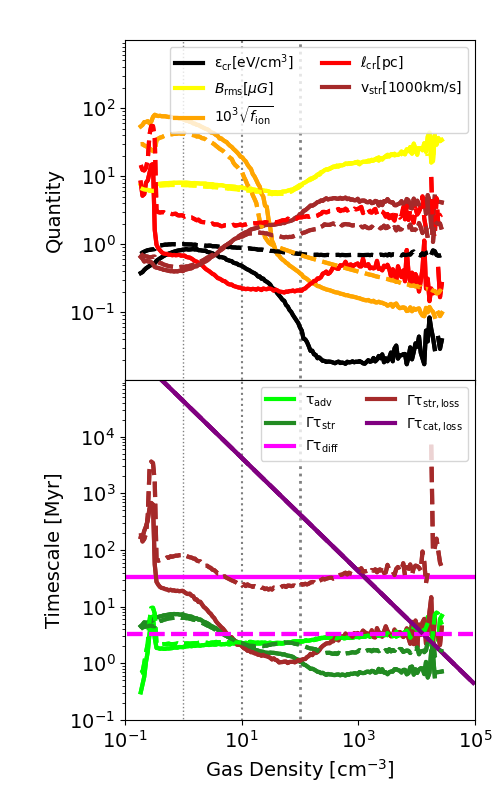}
\caption{Top: Median values of the CR energy density $\epscr$, RMS magnetic field $B_{\rm rms}$, ionization fraction $f_{\rm ion}$, CR gradient scale length $\ell_{\rm cr}=\epscr/\nabla\epscr$, and streaming velocity $\vst$ as a function of density as used in the calculation of the streaming timescale $\tau_{\rm str}$ (Equation \ref{eq:tau_str}) and the streaming loss timescale $\tau_{\rm str, loss}$ (Equation \ref{eq:tau_str_loss}) at $0.5 t_{\rm ff}$. The vertical gray dotted lines correspond to gas densities $n=1, 10, 100 \rm cm^{-3}$. 
Bottom: Advection timescale $\tau_{\rm adv}$, streaming timescale $\tau_{\rm str}$, diffusion timescale $\tau_{\rm diff}$, streaming loss timescale $\tau_{\rm str, loss}$, and collisional loss timescale $\tau_{\rm cat, loss}$ from Equations \ref{eq:tau_adv}, \ref{eq:tau_str}, \ref{eq:tau_diff}, \ref{eq:tau_cat_loss} and \ref{eq:tau_str_loss}, respectively, as a function of gas density for the \fiducial\ (solid) and \highdiffcoeff\ (dashed) simulations at $0.5 \tff$. The transport timescales $\tau_{\rm adv}$, $\tau_{\rm str}$ and $\tau_{\rm diff}$ assume a characteristic length scale of $L=3$ pc. The values of the streaming timescales $\tau_{\rm str}$ and $\tau_{\rm str, loss}$ and the advection timescale $\tau_{\rm adv}$ are calculated using the median values of gas and CR properties at each gas density. All timescales except $\tau_{\rm adv}$ are scaled by $\Gamma_{\rm RSOL}=c/\tilde{c}$. The vertical gray dotted lines correspond to gas densities $n=1, 10, 100 \rm cm^{-3}$.}
\label{fig:timescale_quantities}
\end{figure}

The diffusive transport timescale is inversely proportional to the CR diffusion coefficient. For the \highdiffcoeff\ simulation, diffusive transport is comparable to streaming transport. Physically, the CRs scatter less and so effectively propagate more efficiently. The streaming timescale is also slightly higher due to the higher ionization fraction from the higher CR energy density in the cloud.  Finally and most significantly,  the streaming loss timescale is larger for the \highdiffcoeff\ simulation because the CR distribution is more uniform. The cumulative effect is that in the \highdiffcoeff\ simulation, CRs are able to propagate through the cloud without losing a significant amount of energy. 

Our simulations show that streaming instability energy losses dominate over collisional losses for $\sim$ GeV CRs, which has important physical and observational implications. The CR energy lost to the streaming instability goes into exciting Alfv\`en waves, which are quickly damped by ion-neutral collisions that thermalize the energy. This means that the CR streaming instability provides a stronger source of gas heating then CR ionization.  Furthermore, the CR energy is not lost to $\gamma$-ray production, especially not at the cloud boundary where streaming instability energy losses are strongest. This supports $\gamma$-ray observations from MCs that are in agreement with the solar neighborhood CRIR of $\zeta \approx 10^{-17} \rm s^{-1}$ \citep{krumholz_2023} even though linewidth observations suggest they have elevated CRIRs by approximately an order of magnitude \citep{indriolo_2015}. In other words, our simulations support the conclusion that the high CRIR values observed in Milky Way MCs do not come from an elevated \textit{total} CR background. Instead, they may result from additional sources of low-energy ionizing CRs ($\sim$ MeV), which we discuss more in Section \ref{subsubsection:dis_ionization}. 

On the other hand, $\gamma$-ray measurements from starburst galaxies indicate that they are proton-calorimeters, with all of their CR energy lost to catastrophic collisions. Our results suggest that in these galaxies the streaming speeds in MCs must be slower than in a Milky-Way environment because otherwise the CRs lose all of their energy to streaming losses. The ionization fraction is higher on average in starbursts than in Milky Way type galaxies, with typical values from $\chi \approx 10^{-7}-10^{-4}$ \citep{krumholz_2020}. In regions of higher ionization the ion-Alfv\`en speed is lower and so the CRs can get `trapped' and lose their energy to hadronic losses rather than streaming losses. 

\subsubsection{Exclusion of CRs from Molecular Clouds}
\label{subsubsection:cr_exclusion}

Section \ref{subsubsection:streaming_instability} discusses how streaming instability energy losses cause the CRs initialized in the cloud to quickly lose energy once the cloud starts collapsing. However, this does not explain why CRs from outside the cloud do not propagate into the cloud and offset the energy losses.  In this section, we discuss the effects that may be producing this `exclusion' of CRs from MCs, which has been noted in many previous studies \citep{kulsrud_pearce_1969,skilling_strong_1976,  cesarsky_volk_1978, chandran_2000, padoan_scalo_2005, silsbee_2018, padovani_galli_2011, padovani_hennebelle_galli_2013, everett_zweibel_2011, bustard_zweibel_2021}. 

The primary effect keeping CRs from propagating into the cloud is that streaming energy losses are even stronger at the cloud boundary than inside the cloud. Figure \ref{fig:timescale_quantities} shows that at $0.5 \tff$ the streaming loss timescale of the \fiducial\ simulation drops sharply at the outer cloud boundary region ($n \sim 1-10 \rm cm^{-3}$) and reaches a minimum at the inner cloud boundary region ($n \sim 10-100 \rm cm^{-3}$) because of the sharp gradient in the CR energy density. This indicates that CRs lose a significant amount of energy as they propagate through the boundary and into the cloud. We emphasize that this outcome is not due to the fact that we started the simulations with $\epscr,_{\rm cloud}=0.1\epscr,_{\rm med}$, and it occurs even in tests beginning with $\epscr,_{\rm cloud}=\epscr,_{\rm med}$. Rather, the gradient in the CR energy density always develops at the cloud boundary because of the drop in the ionization fraction and resulting increase in the ion-Alfv\`en velocity there (Figure \ref{fig:timescale_quantities}). 

The degree of CR exclusion from the cloud interior also depends on the diffusion coefficient. Across the entire cloud boundary region ($n \sim 1-100 \rm cm^{-3}$), the diffusive transport timescale is comparable to that of streaming transport.  Physically,  the CRs have a longer path length in the \highdiffcoeff\ simulation, and gradients in the CR energy density are wiped out faster. For the \highdiffcoeff\ simulation, the streaming loss timescale is roughly uniform across the cloud boundary, because no sharp gradients in the CR energy density develop.

\begin{figure*}
\centering
\includegraphics[width=0.98 \textwidth]{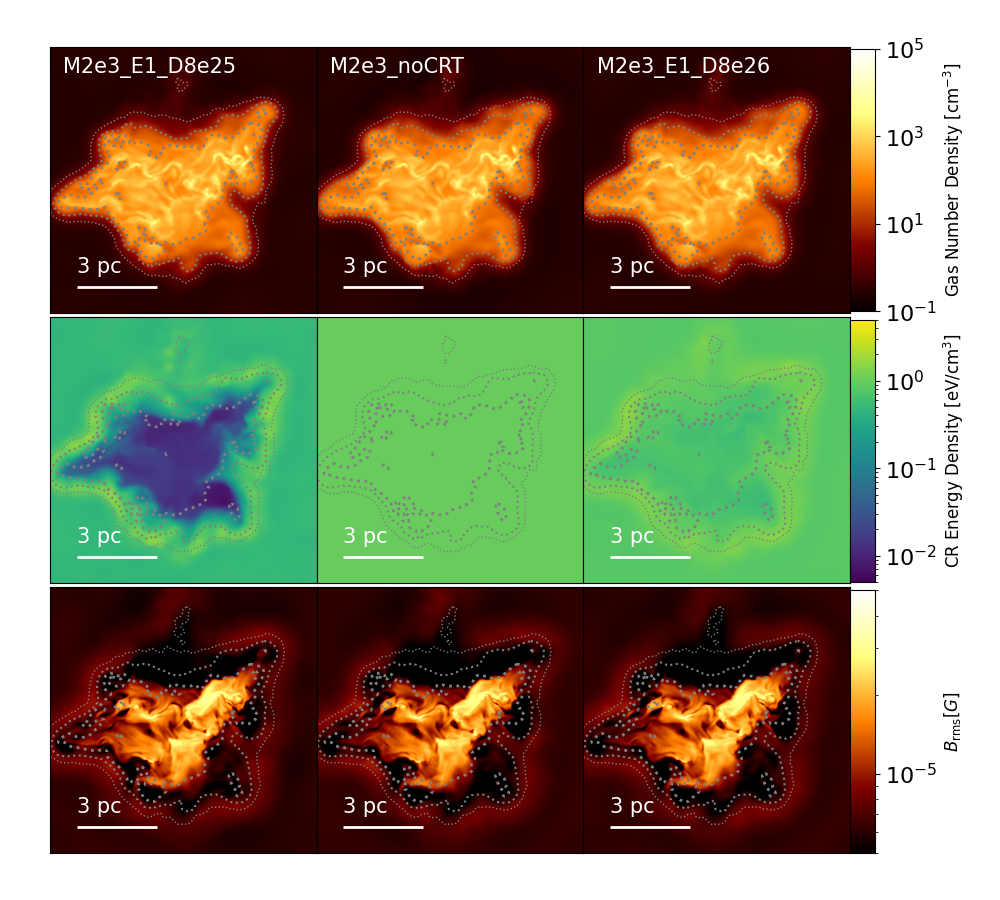}
\caption{Gas number density (top), CR energy density (middle) and RMS magnetic field (bottom) for the \fiducial\ (left), \nocrs\ (middle), and \highdiffcoeff\ (right) simulations at $0.5 \tff$. For the \nocrs\ panel, a CR energy density of $\epscr=1 \rm eV/cm^3$ is shown, corresponding to the assumed ionization rate of $\zeta \sim 1.7 \times 10^{-17} \rm s^{-1}$. The gray dotted contours correspond to gas densities $n=1, 10, 100 \rm cm^{-3}$, also marked in Figure \ref{fig:timescale_quantities}.}
\label{fig:discussion_slice_plots}
\end{figure*}

The CR exclusion at the cloud boundary may also be enhanced by magnetic field effects. Figure \ref{fig:discussion_slice_plots} shows slices of the gas density (top), CR energy density (middle), and root-mean-squared (RMS) magnetic field $B_{\rm rms}$ (bottom) at $0.5 \tff$ through the middle of the domain for the \fiducial\ (left), \nocrs\ (middle) and \highdiffcoeff\ (right) simulations. The magnetic field in the cloud is up to an order of magnitude higher than the magnetic field in the ambient medium, setting up a strong magnetic gradient. However, the magnetic field behavior at the cloud boundary is complex: there is a region with slightly elevated magnetic field that is a few $\mu G$ higher than the medium value at $n \sim 1 \rm cm^{-3}$, followed by a pocket of low-magnetic field that coincides with the boundary region ($n \sim 1-100$cm$^{-3}$). This magnetic behavior is similar in all three simulations and coincides with the phase transition from ionized to neutral gas. In the simulations with CRT, the highly magnetized ring aligns with a peak in the CR energy density (shown by the yellow rim). In the \highdiffcoeff\ simulation, a ring of enhanced CR energy density is faintly visible, but the CR distribution is mostly uniform throughout the cloud and medium. The CRs have a longer path length in the \highdiffcoeff\ simulation, which allows them to more easily scatter over magnetic field asymmetries into the cloud.

We note that in our simulations this `ring' effect may be due to the initialization of the CR energy and the varying cell size across the cloud boundary. If the CRs were able to diffuse freely in this region then the excess may be rapidly smoothed out; instead the enhancement persists even in the \highdiffcoeff\ simulation. Still, the spatial correlation between the CRs and the magnetic field deserves consideration due to the number of studies that have looked at the effects of large-scale non-uniformity of the magnetic field on CR propagation. Magnetic field lines converge in regions undergoing gravitational collapse which results in a proportional increase in the number density of CRs, a phenomenon known as ``focusing". On the other hand, the CRs' pitch angles increase in response to the growing field and more particles are reflected back, known as ``mirroring."  A variety of studies have investigated the competing effects of magnetic mirroring and focusing on CR propagation in MCs \citep{kulsrud_pearce_1969,skilling_strong_1976,  cesarsky_volk_1978, chandran_2000, padoan_scalo_2005, silsbee_2018} and in dense cores \citep{padovani_galli_2011, padovani_hennebelle_galli_2013, fatuzzo_adams_2014}. Which effect dominates depends on the assumed magnetic geometry and details of the CR microphysics. 

Recent studies have suggested that although magnetic fields may partially screen CRs from MCs, the effect is at most modest. \cite{padovani_galli_2011} and  \cite{padovani_hennebelle_galli_2013} find that mirroring dominates over focusing for dense cores but only causes a reduction in the CR energy density by a factor of 2-3 compared to the case where these effects are neglected. \cite{silsbee_2018} found that the two effects cancel except in magnetic ``pockets," local minima in the field strength, where mirroring again dominates by trapping CRs in the pocket; however, even in these regions the reduction in the CRIR is less than an order of magnitude. We conclude that although the CR enhancement along the magnetic field gradient at the boundary may be physical rather than numerical, the primary effect excluding CRs from the cloud interior is streaming instability energy losses.

Past numerical studies have identified CR exclusion from MCs; however, most of them have attributed the cause to effects other than streaming losses.  \cite{skilling_strong_1976} and \cite{cesarsky_volk_1978} found a drop in CR energy density for CRs of $E<300 \rm MeV$ due to ionization losses, thereby creating a CR gradient that amplified the streaming instability at the boundaries of clouds. Although we have a gradient in the CR energy density at the cloud boundary due to the streaming instability, our single-bin approximation of CRs approximates $\sim$ GeV CRs which do not experience the strong ionization losses present in $\sim$ MeV CRs, so this effect is not present here.

Our findings are consistent with a few prior studies. \cite{everett_zweibel_2011} considered a similar simulation configuration in which ionization losses are also subdominant to streaming losses. They showed that even a small CR pressure gradient amplifies the growth of the CR streaming instability. However, they found only a small drop in CRs at the boundaries of clouds rather than the almost complete exclusion we find. \cite{bustard_zweibel_2021} considered more complex cloud geometries and a more advanced CRT model. Their results show gradients in the CR energy density at the cloud boundary similar to those in our simulations. 

Most prior studies \citep{kulsrud_pearce_1969,skilling_strong_1976,  cesarsky_volk_1978, chandran_2000, padoan_scalo_2005, silsbee_2018, padovani_galli_2011, padovani_hennebelle_galli_2013} solve a set of simplified CR propagation equations that model a steady-state, nearly isotropic CR distribution, and they do not model the evolution of the gas and magnetic field. 
Even more importantly, they make the approximation that CR pitch-angle scattering due to the streaming instability is negligible and only model CR streaming along field lines. This is a reasonable assumption on microphysical timescales such as the CR gyro time or ion-neutral damping timescale, where streaming instability energy losses are insignificant \citep{kulsrud_pearce_1969}; however, on the longer timescales of MC evolution comparable to CR transport timescales, CRs are expected to lose a large fraction of their energy  \citep{wentzel_1971}. \cite{everett_zweibel_2011} and \cite{bustard_zweibel_2021} both model CRs interacting with diffuse clouds; thus, they use more advanced moment-based CRT models and consider $\sim$ GeV CRs to more accurately capture streaming transport and scattering. However, \cite{everett_zweibel_2011} do not model the evolution of the cloud properties. \cite{bustard_zweibel_2021} provides the most accurate point of comparison to our simulations; however, their two-moment CRT implementation is only accurate for steady state, close-to-isotropic cases. \textit{None} of these studies considered the full effect of streaming energy losses on the further cloud evolution, which we have shown is an important component of CR physics that should be considered in future studies. 

\subsubsection{Are Streaming Instability Energy Losses Unavoidable in Molecular Clouds?}
\label{subsubsection:si_energy_losses}

The results of our simulations suggest that streaming instability energy losses strongly limit the CR energy density in MCs. In this section, we discuss whether this is an unavoidable outcome of CR physics in MCs. 

We first consider the situation where the CRT is streaming dominated, which is the case in all except the \highdiffcoeff\ simulation. In this limit, the Eulerian equation for the CR energy density solved by many other studies \citep[e.g.,][]{chan_2019} converges to $\partial \epscr/\partial t = -\epscr \nabla \cdot \bold{\vst} - \gammacr|\vst| \epscr/\ell_{\rm cr}$. This equation limits to $\epscr \rightarrow 0$ if the streaming velocity increases going into the cloud ($\nabla \cdot \bold{\vst}$ is positive) and/or  $|\epscr \nabla \cdot \bold{\vst}| <|\gammacr\vst\epscr/\ell_{\rm cr}|$. The only way for $\epscr$ to stabilize at a higher value is if $|\epscr \nabla \cdot \bold{\vst}| > |\gammacr\vst\epscr/\ell_{\rm cr}|$ \textit{and} $\vst$ decreases going into the cloud. However, this is opposite of the expected behavior, because the ionization fraction decreases in the cloud and the streaming velocity increases (Figure \ref{fig:timescale_quantities}). Consequently, streaming losses are unavoidable if streaming is the dominant transport mechanism.

Next we consider the case where {streaming and diffusive transport are comparable} as in our \highdiffcoeff\ simulation. 
This simulation models an environment with a strong macroscopic diffusivity $D_{\parallel, \rm FLW}$ where turbulence advects the magnetic field lines at a rate faster than streaming energy losses remove energy.  However, this scenario was only realized by ad-hoc setting the diffusion coefficient to a high value and is not consistent with the degree of field-line wandering expected in mostly-neutral gas unless $f_{\rm ion} \lesssim 10^{-9}$ \citep{sampson_2022}. Figure \ref{fig:ionization_fraction_phase} shows that an ionization fraction this low is rarely achieved in our simulations, especially near the cloud boundary, and, furthermore, is not in line with measured values in MCs, \citep{caselli_1998}. Moreover, this limit could not be sustained because the CRs would stream with minimal energy losses, which would increase $f_{\rm ion}$ and thereby decrease the ion Alfv\'en velocity. 

It is also impossible to achieve this limit with microphysical diffusion. As discussed in Section \ref{subsubsection:variable_cr_diffusion_coefficient}, if the streaming instability is the only source of microphysical scattering then achieving $v_{\rm diff} \gg \vAi$ would require $\epscr \ll 1 \rm eV/cm^{3}$. Adding additional sources of CR scattering from extrinsic turbulence would decrease $D_{\parallel, \rm micro}$; in other words, assuming scattering due to self-generated turbulence alone provides an upper limit to $D_{\parallel, \rm micro}$. Therefore, it is not possible to reach the diffusion-dominated transport limit unless $\epscr \ll 1 \rm eV/cm^{3}$ in the cloud from the start. 

An alternative way to reduce streaming losses is to suppress the growth of the streaming instability by limiting the drift velocity to less than the ion-Alfv\`en velocity ($v_{D} = \vst+v_{\rm diff} < \vAi$). In steady state, the growth rate of the streaming instability (Equation \ref{eq:streaming_instability_growth_rate}) is matched by the ion neutral damping rate (Equation \ref{eq:ion_neutral_damping_rate}), which sets the diffusive component of the velocity $v_{\rm diff}$. However, Equation \ref{eq:streaming_instability_growth_rate} shows that when $v_D < \vAi$, $\Gamma_{\rm SI} < 0$ and the streaming instability will never grow. In our simulations, we directly model CR streaming using $\vst =\vAi$ and include $v_{\rm diff}$ using the diffusion coefficient $D_{\parallel}$, so by construction we cannot capture a situation where $v_{D} < \vAi$. However, in principle the streaming velocity could be less than the ion Alfv\`en velocity. One way to achieve this is if $\vAi \gtrsim c/3$ (where the factor of 3 comes from appropriately including the moments of the CR distribution function). Using appropriate gas quantities, ($B \sim 10 \mu G, n \sim 100 \rm cm^{-3}, m_{\rm ion} \sim 29 m_p$), $\vAi \approx 0.5 f_{\rm ion}^{-1/2} \rm km/s$ and so achieving $\vAi \gtrsim c/3$ would require $f_{\rm ion} \lesssim 10^{-9}$. For the reasons discussed above, this limit could not be sustained. 

Another way to keep $v_{D} < \vAi$ is through an extrinsic source of strong, symmetric scattering modes that scatter the CRs isotropically. The full expression for the CR streaming speed is $\vst = \vAi (\barnu^{+} -\barnu^{-})/(\barnu^{+} +\barnu^{-})$, 
where $\barnu^+$ and $\barnu^-$ are the scattering modes for forward and backward propagating waves, respectively \citep{hopkins_2021a}. Extrinsic sources of turbulence tend to produce symmetric scattering with $\barnu^+ \approx \barnu^-$. In this limit, $\vst \rightarrow 0$ so the $\vst$ terms in Equations \ref{eq:cr_energy_one} and \ref{eq:cr_energy_two} vanish and the diffusive acceleration term $\Ssc$ dominates, meaning the CRs may gain energy as they propagate. 

For this model to be viable, the turbulence must be strong enough such that $v_{\rm diff} \sim D_{\parallel, \rm micro}/\ell_{\rm cr} \ll \vAi$ so the streaming instability is not excited in the first place. Using appropriate quantities for the diffuse ISM ($B \sim \mu G, n \sim 1 \rm cm^{-3}, m_{\rm ion} \sim m_p$), $\vAi \approx 2 \rm km/s$  $f_{\rm ion}^{-1/2}$, and so $D_{\parallel} \lesssim 5 \times 10^{23} (\ell_{\rm cr}/\rm pc$) $f_{\rm ion}^{-1/2} \rm cm^2/s$. This value is much lower than the microphysical diffusivity in the diffuse ionized ISM ($f_{\rm ion} \approx 1$), where simulations estimate the diffusivity at a few $\times 10^{28} \rm cm^2/s$ \citep{trotta_2011}. Moreover, this scattering would have to be strong enough to compete with ion-neutral damping in MCs. We tested this scenario by including a standard model for extrinsic turbulence scattering in our \vardiffcoeff\ simulation and found that this limit was never reached (otherwise, there would have been no streaming transport or losses in that simulation).

We conclude that any scenario in which streaming instability energy losses are avoided in MCs requires new physics. Either our understanding of the CR streaming instability and self confinement is wrong, or there are sources of extrinsic turbulence that are stronger than any currently known that can produce the necessary scattering. 

\subsection{Comparison to Observations}
\label{subsection:observations}

\subsubsection{Ionization Within Molecular Clouds}
\label{subsubsection:dis_ionization}

Measured values of the CRIR in the Milky Way vary by over an order of magnitude. The value measured directly by the \textit{Voyager} spacecraft, which has the lowest amount of uncertainty, is $\zeta \sim 2 \times 10^{-17} \rm s^{-1}$ \citep{stone_2019}. However, this only probes the CRIR in the Solar neighborhood. Indirect measurements of the CRIR obtained from $\rm H_3^+$ observations of diffuse molecular gas suggest the CRIR varies over a few orders of magnitude within the Milky Way, with an average value of a few $\times 10^{-16} \rm s^{-1}$ \citep{indriolo_2012, indriolo_2015, neufeld_wolfire_2017, kalosi_2023}. 

Although most of our simulations assume a CR background consistent with that measured by \textit{Voyager}, only the \highdiffcoeff\ simulation reached a CR level throughout the cloud of $\zeta \sim 10^{-17}$. Our \highcrs\ simulation, which adopted an external CR energy density 10 times that measured by \textit{Voyager}, had a larger range of CRIRs throughout the cloud than the other simulations (top right panel of Figure \ref{fig:cr_energy_density_phase}). However, it still had a significantly lower CR energy density inside the cloud due to CR exclusion. These results suggest that the observed elevated ionization rates are unlikely to come from spatial variations in the galactic CR distribution pervading the clouds. For low to moderate values of the CR diffusion coefficient, the CRs lose a significant amount of energy due to the streaming instability. For high values of the CR diffusion coefficient, the CRs do not scatter enough to create significant spatial variations.

Measurements of higher density gas tracers indicate that there is tentative evidence that the CRIR declines in higher column density gas \citep[e.g.,][]{padovani_2023}. Ionization maps of dense MCs using $\rm H_2D^+$ show CRIRs lower than those in diffuse clouds ranging from a few $\times 10^{-17} - 10^{-16} \rm s^{-1}$, in agreement with this conclusion \citep{sabatini_2023}. However, they also show that within MCs the CRIR often spans orders of magnitude as well. Observations of $\t{HCO}^+$ and $\t{N}_2\t{H}^+$ along sightlines towards the intermediate-mass protostar OMC-2 FIR 4 indicate values of  $\approx 10^{-14} \rm s^{-1}$, a couple orders of magnitude {\it higher} than the mean CRIR \citep{ceccarelli_2014, fontani_2017}. Meanwhile, ionization maps of the Class 0 protostar B335 show large fractions of ionized gas ($f_{\rm ion} \approx 1-8 \times 10^{-6}$) and suggest the ionization increases towards small envelope radii, reaching values of $\sim~ 10^{-14} \rm s^{-1}$ a few hundred au from the central protostar \citep{cabedo_2023}. 

Models used to derive the CRIR using dense gas tracers have significant uncertainties and their accuracy varies with the source properties \citep{shingledecker_2016}. Still, none of our simulations produced regions approaching the elevated values inferred for B335. They show that streaming instability energy losses result in a uniform CRIR throughout the cloud, without enhancements in denser regions. 

Our study lends support to the idea that there may be internal sources of CR acceleration within MCs associated with the protostellar shocks \citep{padovani_2016, gaches_2018, fitzaxen_2021}. \cite{krumholz_2023} estimate that protostellar sources contribute $\sim 1/4-1/3$ as much energy to the galactic CR energy budget as stellar winds. Low mass stars form earlier and are more numerous than high mass stars, and so would accelerate CRs earlier in the cloud collapse than stellar winds. These CRs may explain the observed high CRIRs, especially near protostars, as well as the overall elevated ionization rates 
inferred in diffuse clouds. However, the degree of ionization inside clouds is sensitive to CR physics and the CR diffusion coefficient, which is highly uncertain. Thus, ionization rate measurements and comparisons based on synthetic observations can provide helpful constraints for models of CR acceleration and propagation within MCs. 

\subsubsection{Star Formation in a High CR environment}
\label{subsubsection:high_cr_environment}

Many regions in the Milky Way and other galaxies have a CRIR that is higher than the \textit{Voyager} value of $\sim 10^{-17} \rm s^{-1}$. In particular, recent observations show evidence of elevated CRIRs towards the center of galaxies. Although the Galactic average is $\approx 2 \times 10^{-16} \rm s^{-1}$ \citep{indriolo_2015}, the CRIR measured in the central parsec of the Milky Way is $\approx 2$ orders of magnitude higher than that in the clouds outside the galactic center \citep{goto_2014, indriolo_2015, rivilla_2022}. Extragalactic measurements towards the central molecular zone of the starburst galaxy NGC 253 suggest even higher values. \cite{holdship_2022} studied several star-forming regions there and found CRIR values between $1-80 \times 10^{-14} \rm s^{-1}$. \cite{behrens_2022} found that the central GMCs of NGC 253 had $\zeta \approx 10^{-13} - 10^{-11} \rm s^{-1}$, greater than four orders of magnitude higher than the Milky Way value.

We note that the accuracy of these measurements are in tension with theoretical predictions. \cite{krumholz_2023} argue that these CRIR values are inconsistent with the measured star formation rates, which determine the amount of energy available for accelerating CRs and that the true CRIRs are significantly lower, though still likely a factor of 10-100 greater than the \textit{Voyager} value. They are also inconsistent with other observational and theoretical expectations, which predict an anticorrelation between CRIR and gas density \citep{indriolo_2012, neufeld_wolfire_2017, padovani_2023}. However, regions of the ISM near SNRs may have elevated CR levels, as SNRs are thought to be the primary accelerators of Galactic CRs \citep{aharonian_2013}. Indeed, measurements of CRIRs targeting clouds near SNRs  show elevated rates of $\approx 100$ times the Galactic average \citep{ceccarelli_2011, vaupre_2014}. 

We found that star formation occurs more rapidly and the SFE is a few percent higher  in our \highcrs\ simulation, which raises the CR background level by only an order of magnitude. Thus, we might expect larger differences in star formation in even more extreme CR environments, such as near SNRs. However, the details of the effect are highly sensitive to the assumed value of the diffusion coefficient. Raising the diffusion coefficient by an order of magnitude caused a large difference in the CR propagation within the cloud, diminishing energy losses from the streaming instability and exclusion by magnetic fields. Meanwhile, decreasing the diffusion coefficient by an order of magnitude caused even more rapid CR energy losses. These results suggest a higher diffusion coefficient and a higher CR environment acting in concert could produce more extreme variations in star formation outcomes. Further work is needed to investigate this parameter space. 

\subsection{Simulation Caveats}
\label{subsection:simulation_caveats}

\subsubsection{Energy-Independent Treatment of CRs}
\label{subsubsection:caveats_one}

Our simulations make a number of approximations in the treatment of CR physics. We use a single-bin model for the CR fluid that represents the total CR energy rather than evolving a full multi-bin energy spectrum. The treatment of the CR velocities and the energy loss function assume $\sim$ GeV CRs.  Simulations modeling a full multi-bin CR energy spectrum that extends down to $\sim$ MeV and up to $\sim$ TeV energies would likely predict different CR behavior within the dense gas. For example, $\sim$ MeV CRs more efficiently ionize gas than $\sim$ GeV CRs. Collisional energy losses are anticorrelated with gas density (see Equation \ref{eq:cr_energy_loss}), so multi-bin CRT simulations might better reproduce the apparent density-dependent CRIR seen in observations \citep{padovani_2023}. Modeling CRs with $\gtrsim$ GeV energies would likely have a minimal impact on our results, because these CRs are less numerous than CRs of lower energies \citep{stone_2019}. However, these CRs produce $\gamma$-rays, so comparisons to observed $\gamma$-ray emission could lead to additional constraints on CRT in MCs.  

We note that modeling a full CR spectrum would not change the conclusions presented in \ref{subsubsection:si_energy_losses} because they apply at all CR energies as long as the streaming instability dominates the CR scattering. CRs of all energies will stream at the ion Alfv\`en speed and experience streaming energy losses. Thus, the cloud will still develop a low CR energy density eventually, even if there is measurable ionization and $\gamma$ ray emission before the CRs lose their energy.

\subsubsection{Reduced Speed of Light Approximation}
\label{subsubsection:caveats_two}

We use a reduced-speed-of-light (RSOL) of 300 km/s for our simulations to make them computationally tractable. This is higher than the value used for other STARFORGE calculations of $\tilde{c} = 30 \rm km/s$ \citep{grudic_2021, grudic_2022} or $\tilde{c} = 90 \rm km/s$ \citep{grudic_2023}, which has been thoroughly tested for non-CRT simulations in which the RSOL affects the radiative transfer calculation. The choice of RSOL should not have a significant impact on our results as long as it is higher than the other relevant velocities in the simulations such as the CR streaming speed $\vst$ and the diffusion speed set by the diffusivity $D_{\parallel}$.  Otherwise, the CR transport and loss processes may be artificially slowed down \citep{hopkins_2020}. As discussed in Section \ref{subsubsection:streaming_instability} and shown in Figure \ref{fig:discussion_slice_plots}, the streaming velocity is an order of magnitude higher than 300 km/s in many regions of the cloud. In order to justify our choice of RSOL, we run a suite of low-resolution simulations that vary the RSOL. We find that using a higher RSOL increases the speed of CR transport and rate of CR energy losses in the cloud but does not significantly effect the star formation properties. We show these results and discuss the implications further in Appendix \ref{section:rsol_appendix}. 

\subsubsection{Sub-Grid Models and Resolution}
\label{subsubsection:caveats_three}

Our simulations predict an IMF with a high-mass slope $\alpha=-1$ consistent with results from previous STARFORGE calculations (Figure \ref{fig:imf}). As discussed in \cite{grudic_2022}, limitations due to the feedback models and resolution may produce a shallower IMF than the commonly assumed value of $\alpha=-1.35$. 

The STARFORGE protostellar jet feedback model, discussed in Section \ref{subsubsection:starforge_physics}, uses  fixed parameters $f_w=f_K=0.3$, which parameterize the jet velocities but are observationally uncertain \citep{cunningham_2011}. Variations in these parameters which increase the jet velocity may regulate massive SF
\citep{guszejnov_2021, rosen_2021}. 

We are limited in resolution in both length scale and mass \citep{grudic_2022, guszejnov_2022}.  Size scales smaller than the Jeans length are not well resolved in the simulations, so protostellar disks are unresolved. Disk fragmentation may affect the stellar multiplicity and mass ratio distribution of binaries \citep{offner_2023}, so fully resolving disks may steepen the IMF.  

\subsubsection{Calculation of the Streaming Speed}
\label{subsubsection:caveats_four}

To calculate the ion Alfv\`en velocity, {\small GIZMO} assumes $m_{\rm ion}=m_p$, which is equivalent to setting $\chi=f_{\rm ion}$. This is not a correct assumption for MCs, where the dominant charge carrier is $\rm HCO^+$ ($m_{\rm ion}=29 m_p$), and so overestimates $\vAi$ by a factor of $\sim \sqrt{29} \approx 5$. Additionally, we do not account for the effect of dust damping on the CR streaming speed. Dust grains can damp the amplitude of the Alfv\`en waves from the streaming instability, reducing the scattering rate and increasing the streaming speed \citep{squire_2021}.

Since both streaming transport and losses scale with $\vAi$, these caveats do not change the major conclusions presented throughout Section \ref{section:discussion} emphasizing the importance of streaming losses when the CR transport is dominated by streaming. However, they may modify when CRs are in the streaming versus diffusion dominated regimes (Section \ref{subsubsection:streaming_instability}). These uncertainties are comparable to others in the chemistry module, and for given values of $\vst$ and $v_{\rm diff}$ our conclusions are robust. 

\section{Conclusions}
\label{section:conclusions}

In this work, we present numerical simulations of the collapse of MCs including CRT as part of the STARFORGE project. These simulations follow the formation of individual stars and include all major stellar feedback mechanisms including protostellar jets, radiation, and stellar winds. We expand on previous STARFORGE simulations by explicitly modeling CRT using a single-energy bin model of CRs, which are evolved as a fluid using an M1-moment based method. We explore variations in the initial CR energy density and CR diffusion model. Our main findings are as follows:

1. In a fiducial CR environment where $\epscr = 1 \rm eV/cm^3$, including CRT does not produce a strong difference in the dynamics of the gas collapse compared to simulations without CRT (Figures \ref{fig:proj_plots_fid} and \ref{fig:proj_plots_noCRs}). This is true regardless of the model used for the diffusion coefficient (Figure \ref{fig:simulation_proj_comparison}). 

2. In a high CR environment where $\epscr = 10 \rm eV/cm^3$, the enhanced CR pressure outside the cloud causes faster collapse, earlier star formation and a higher final SFE (Figure \ref{fig:simulation_proj_comparison}). The SFE in the high CR environment increases earlier due to the earlier cloud collapse but progresses at the same rate as clouds embedded in a typical CR environment.

3. For low to moderate values of the CR diffusion coefficient, streaming instability energy losses strongly attenuate the CRs (Figure \ref{fig:fid_cr_energy_density}) . However, when the diffusion coefficient is high enough ($\gtrsim 10^{26} \rm cm^2/s$) the CRs lose less energy and distribute uniformly throughout the cloud at a level similar to that outside the cloud (Figure \ref{fig:simulation_proj_comparison}). 

4. In a CR environment where $\epscr = 1 \rm eV/cm^3$ the evolution of the SFE is insensitive to the CR treatment (Figure \ref{fig:sfe_mmed}). The final SFE is slightly higher for all of the CRT clouds than for the cloud with no explicit CRT. The no-CRT simulation has a $\rm SFE \approx 7$ \%, while all of the CRT simulations have $\rm SFE = 8-10$ \%. The high-CR cloud has a $\rm SFE> 12$ \%. 

5. Variations in the CR diffusion coefficient do not have a statistically significant impact on the IMF of the clouds (Figure \ref{fig:imf}). All simulations in an environment with $\epscr = 1 \rm eV/cm^3$ have a final stellar mass distribution consistent with previous STARFORGE calculations, with a peak at $\sim 0.2-0.3 M_{\odot}$ and a high mass slope of $\alpha \approx -1$.  However, one of the CRT simulations (\fiducial) has a higher median mass, possibly due to stellar wind CR feedback that changes the gas conditions inside the cloud. In a high-CR environment with $\epscr = 10 \rm eV/cm^3$ the cloud forms both more low mass ($M< 0.1 M_{\odot}$) and high mass ($M> 1 M_{\odot}$) stars. 

6. In a fiducial CR environment the CRIR is roughly uniform throughout the cloud (Figure \ref{fig:cr_energy_density_phase}). For low to moderate values of the CR diffusion coefficient it is up to a few orders of magnitude less than outside the cloud, with a median value $ \zeta \lesssim  2 \times 10^{-19} \rm s^{-1}$ at $1-1.5 \tff$. In a high-CR environment simulation the CRIR in the cloud is higher than in a fiducial environment ($\zeta \lesssim 6 \times 10^{-18} \rm s^{-1}$) and there is a larger spread of values, but is still less than the commonly assumed value of $\sim 10^{-17} \rm s^{-1}$ throughout most of the cloud. Overall, the CR background does not provide a significant contribution to the ionization of the cloud compared to radiative feedback once star formation is underway (Figure \ref{fig:ionization_fraction_phase}).

Further numerical studies exploring the impact of a high CR diffusion coefficient in high-CR environments may provide further insights and constraints on CRT. Furthermore, numerical simulations including locally injected CRs from protostellar sources may explain the elevated CRIRs and variations within clouds observed from various sources.  

\begin{acknowledgements}

This research is part of the Frontera computing project at the Texas Advanced Computing Center, and used computing award AST21002. Frontera is made possible by National Science Foundation award OAC-1818253. This research was supported in part by NASA ATP grant 80NSSC20K0507.  This material is based upon work supported by the U.S. Department of
Energy, Office of Science, Office of Advanced Scientific Computing Research, Department of
Energy Computational Science Graduate Fellowship under Award No. DE-SC0021110. Support for PFH was provided by NSF Research Grants 1911233, 20009234, 2108318, NSF CAREER grant 1455342, NASA grants 80NSSC18K0562, HST-AR-15800. MRK acknowledges support from the Australian Research Council with the \textit{Discovery Projects} and \textit{Laureate Fellowship} schemes, awards DP 230101055 and FL220100020. Support for MYG was provided by NASA through the NASA Hubble Fellowship grant \#HST-HF2-51479 awarded  by  the  Space  Telescope  Science  Institute,  which  is  operated  by  the   Association  of  Universities  for  Research  in  Astronomy,  Inc.,  for  NASA,  under  contract NAS5-26555.  

\end{acknowledgements}

\appendix

\section{Choice of the Reduced Speed of Light}
\label{section:rsol_appendix}

We use a RSOL of $\tilde{c} = 300$ km/s for both the CRs and radiation in our simulations. This makes the simulations computationally feasible by limiting the timestep criterion to $\Delta t \leq C_{\rm cour}\Delta x/\tilde{c}$ in all regions of the grid. The choice of $\tilde{c}$ should be irrelevant if it is much faster than the other velocities in the simulation, as the CR propagation equations converge to the same solutions in steady-state independent of $\tilde{c}$. However, the top panel of Figure \ref{fig:timescale_quantities} shows that in many regions during the cloud evolution the CR streaming velocity is larger than 300 km/s. Consequently, all transport and loss timescales except the advection timescale are slowed down by a factor of $\Gamma_{\rm RSOL}=c/\tilde{c} = 1000$ \citep{hopkins_2020}, which is used to scale the timescales plotted in the bottom panel of Figure \ref{fig:timescale_quantities}.

It is not feasible to run the simulations at full resolution with $\tilde{c} \sim 10^3-10^4 \rm km/s$ in a reasonable amount of wall clock time. 
Instead, we test the impact of our choice of $\tilde{c}$ by performing six low resolution simulations ($N_{\rm gas} = 2 \times 10^5$) that vary the value of $\tilde{c}$. Three of these start with the initial CR background equivalent to that used in the \fiducial\ simulation ($\epscr, _{ \rm med}$ = 1 $\rm eV/cm^3$) and three  start with the initial CR energy density used in the \highcrs\ simulation ($\epscr, _{ \rm med}$ = 10 $\rm eV/cm^3$). For each initial CR configuration, we adopt values $\tilde{c} = 75, 300$, and 1200 km/s. All six tests use the fiducial diffusion coefficient $D_{\parallel} = 8.33 \times 10^{25} \rm cm^2/s$. The six different runs are listed in Table \ref{table:table_appendix}.

\begin{figure*}
\centering
\includegraphics[width=0.99 \textwidth]{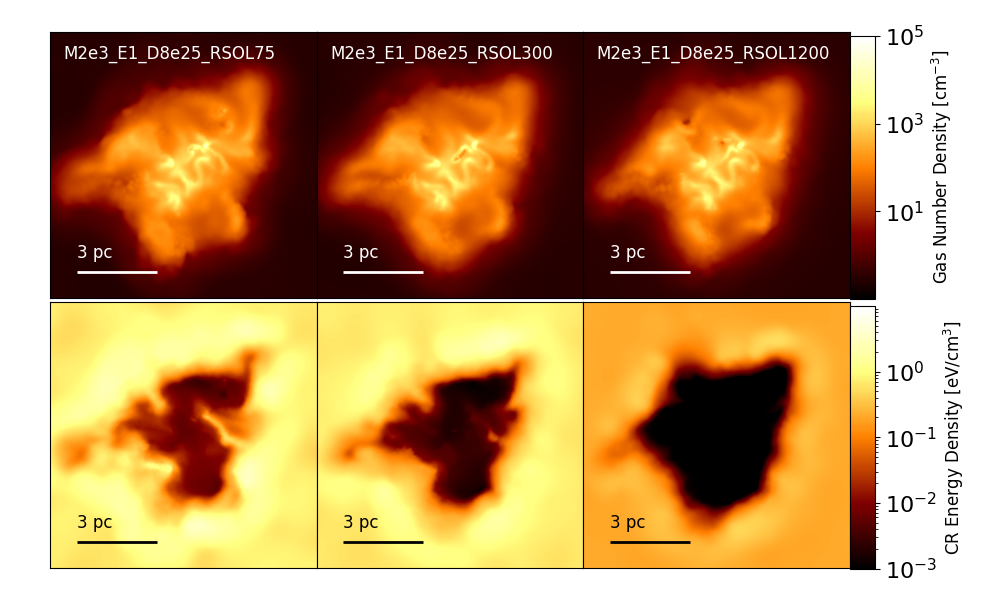}
\caption{Gas number density (top) and CR energy density (bottom) at $1\tff$ for the runs with a reduced-speed-of-light $\tilde{c}=75$ km/s (left), $\tilde{c}=300$ km/s (middle), and $\tilde{c}=1200$ km/s (right), starting with the \fiducial\ CR distribution.}
\label{fig:rsol_slices}
\end{figure*}

\begin{figure}[th!]
\centering
\includegraphics[width=0.48 \textwidth]{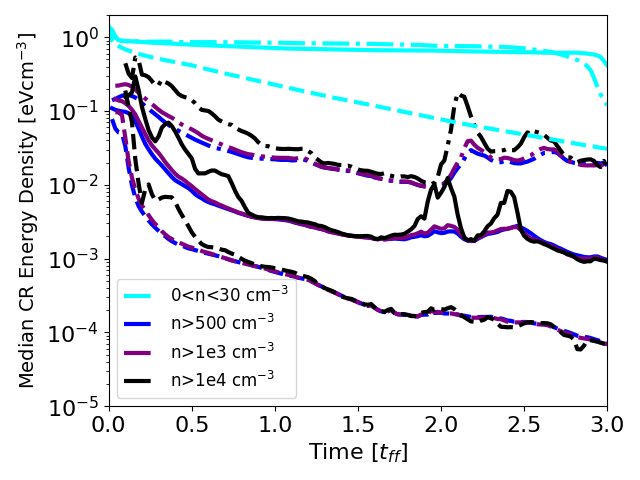}
\includegraphics[width=0.48 \textwidth]{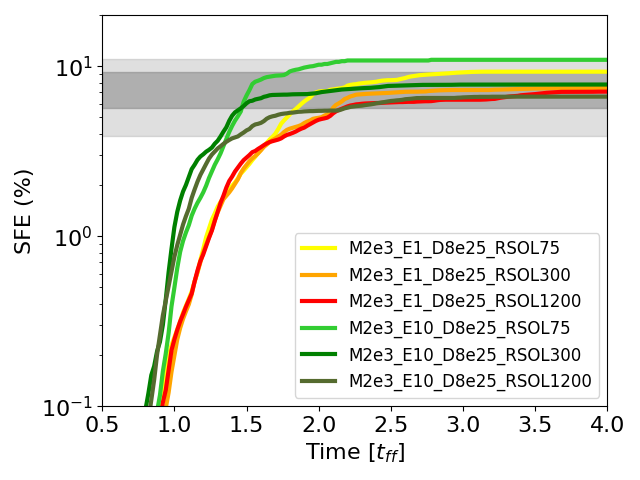}
\includegraphics[width=0.48 \textwidth]{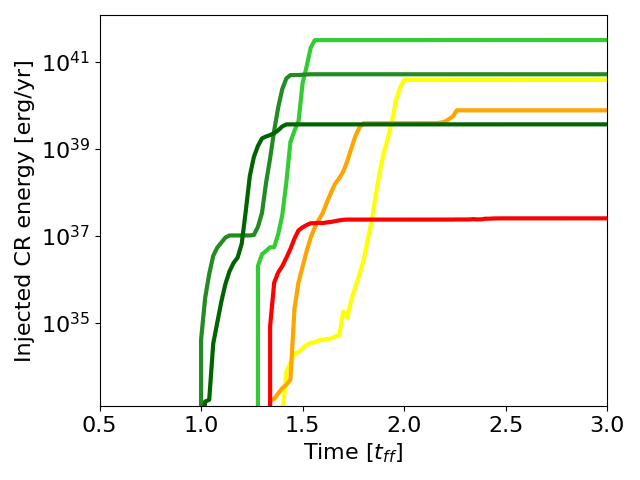}
\includegraphics[width=0.48 \textwidth]{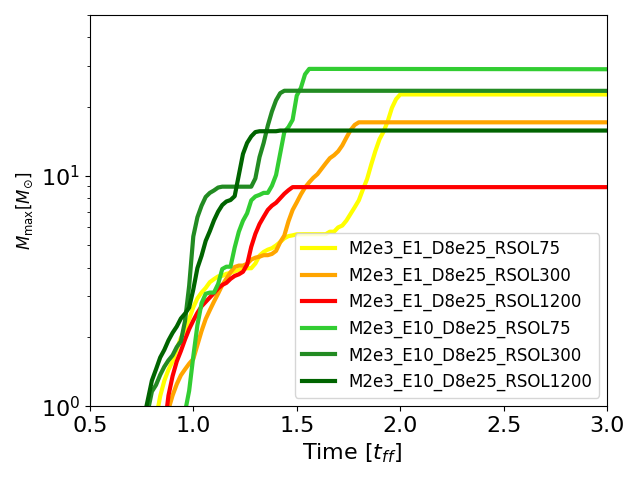}
\caption{Top Left: Median CR energy density for gas in the indicated density range or lower limit for the M2e3\_E1\_D8e25\_RSOL75 (dashdot), M2e3\_E1\_D8e25\_RSOL300 (solid), and M2e3\_E1\_D8e25\_RSOL1200 (dashed) simulations. Top Right: Evolution of the SFE for the RSOL tests. Shaded regions are the $1\sigma$ (dark gray) and $2\sigma$ (light gray) final SFE values obtained from a set of 100 2,000 M$_\odot$ cloud simulations without CRT that vary the initial turbulent seed. 
We note that both the tests with the fiducial RSOL of $\tilde{c}=300 \rm km/s$ do not evolve exactly identically to the full resolution analogs due to numerical differences. Bottom left: Evolution of the total CR energy injected by stellar winds for the RSOL tests. Bottom right: Evolution of the maximum stellar mass for the RSOL tests.}
\label{fig:rsol_evolution_plots}
\end{figure}

Figure \ref{fig:rsol_slices} shows slice plots of the gas number density (top) and CR energy density (bottom) for the \fiducial\ setup at $t=\tff$ using $\tilde{c} = 75 \rm km/s$ (left), $\tilde{c} = 300 \rm km/s$ (middle), and $\tilde{c} = 1200 \rm km/s$ (right). Increasing $\tilde{c}$ causes faster CR energy losses because the streaming loss timescale $\tau_{\rm str, loss}$ scales inversely with $\tilde{c}$. The biggest discrepancy occurs near the cloud boundaries where streaming energy losses are significant, as discussed in Section \ref{subsubsection:cr_exclusion}. The top left panel of Figure \ref{fig:rsol_evolution_plots} shows the evolution of the median CR energy density as a function of time for various gas densities for the three \fiducial\ setup tests. This plot shows that by $1\tff$ the $\tilde{c} = 75 \rm km/s$ run has $\epscr\gtrsim 10^{-2} \rm eV/cm^3$ in the cloud, while the $\tilde{c} = 1200 \rm km/s$ run has $\epscr < 10^{-3} \rm eV/cm^3$. The variation between different choices of $\tilde{c}$ is slightly more than the factor of four difference in $\Gamma_{\rm RSOL}$, which indicates that the choice of $\tilde{c}$ has an impact on the CR energy evolution even greater than the difference expected from timescale analysis. 

Another potential consequence of using a low value of $\tilde{c}$ is that CRs can become artificially trapped in dense gas \citep{hopkins_2020}. Since the calculations adopt a Lagrangian frame of reference that moves with the gas, the advection timescale $\tau_{\rm adv}$ is independent of $\tilde{c}$, but the streaming and diffusion timescales $\tau_{\rm str}$ and $\tau_{\rm diff}$ are not. Using a low value of $\tilde{c}$ can cause CR advection to dominate and not allow the CRs to escape through diffusion and streaming. Figure \ref{fig:rsol_evolution_plots} shows that all three tests have slightly elevated values in gas densities $n> 10^4 \rm cm^{-3}$ (black lines) at early times but reach a uniform CR energy density in the cloud at slightly earlier times for a higher value of $\tilde{c}$. Still, this difference is minor compared to the discrepancy between the different tests for the same density range, indicating that the enhancement caused by artificial CR trapping is subdominant compared to the differences caused by streaming energy losses. However, this effect may produce more significant differences in the evolution for smaller clouds.

In addition to these differences in the CR transport, there is a correlation between the value of $\tilde{c}$ used and the stellar mass distribution formed. Figure \ref{fig:rsol_evolution_plots} shows that the M2e3\_E1\_D8e25\_RSOL1200 simulation does not have a noticeable increase in the median CR energy density between $\sim 1.5-2.5 \tff$ like the other \fiducial\ setup tests. The stars formed in that test simulation did not have stellar winds powerful enough to inject a significant amount of CR energy. To illustrate this point, the bottom row of  Figure \ref{fig:rsol_evolution_plots} shows a plot of the evolution of the CR energy injected by stellar winds (left) and the most massive star in the simulation (right) for our six test runs. For a given initial CR distribution, a higher value of $\tilde{c}$ produces more rapid initial accretion onto a few massive stars, causing the injected CR energy to rise as the stellar winds become more powerful due to the more massive stars. However, this initial accretion phase is cut off earlier, capping the stellar wind energy. In the M2e3\_E1\_D8e25\_RSOL1200 simulation, no stars form that are massive enough to have stellar winds that noticeably increase the median CR energy density in the cloud. 

Despite these caveats, Figure \ref{fig:rsol_slices} shows that the evolution of the gas density is similar for all choices of $\tilde{c}$. This is partially because the \fiducial\ setup is initialized with $ \epscr = 1 \rm eV/cm^3$, which is not high enough to strongly influence the gas dynamics even with minimal energy losses. Additionally, the effects described on the CRT are not as drastic as they initially seem because at the relevant length scales our choice of $\tilde{c}$ is large enough. Figure \ref{fig:timescale_quantities} shows the timescales for CRs to propagate all the way through the cloud ($L=3$ pc) through streaming and diffusion \textit{alone} are $\approx 1-10$ Myr, comparable to the advection timescale (which is a good approximation of the dynamical timescale). However, since the advection timescale is independent of $\tilde{c}$, the gas collapse transports CRs through the cloud and helps offset the slow transport timescales. The CR energy density will reach a relatively smooth distribution on smaller timescales than shown in Figure \ref{fig:timescale_quantities} because the simulation resolution is much smaller than the cloud radius ($\Delta x_J \sim 36 \rm au$), and CR gradients at these scales will be wiped out quickly. Figures \ref{fig:fid_cr_energy_density} and \ref{fig:timescale_quantities} show that by $0.5-0.75 \tff$ the CR energy density at $n\gtrsim 500 \rm cm^{-3}$ is roughly independent of gas density, indicating that CRs are indeed able to propagate out of the densest gas. 

Furthermore, despite the differences in the stellar accretion, differences in the evolution of the CR distribution in the cloud do not change the SFE until near the end of the cloud evolution.  Figure \ref{fig:rsol_evolution_plots}, top right,  shows the evolution of the SFE for the six different runs. As in Section \ref{subsection:star_formation}, we assume SFE $\approx 7.42 \pm 1.77$ as the $1\sigma$ region, although the exact numbers should not be compared to our main simulations with $N_{\rm gas} = 2 \times 10^6$ as resolution causes variations in the results as well. For both the \fiducial\ variations and \highcrs\ variations, the evolution of the SFE begins similarly for the different values of $\tilde{c}$. However, the time at which star formation slows and the cloud disperses occurs later for a lower choice of $\tilde{c}$, raising the final value of the SFE. 

The final values of the SFE for the six runs are shown in Table \ref{table:table_appendix}. If we assume SFE $\approx 7.42 \pm 1.77$, then both the M2e3\_E1\_D8e25\_RSOL75 and M2e3\_E10\_D8e25\_RSOL75 tests have values of the SFE above the $1\sigma$ region. The difference between the M2e3\_E1\_D8e25\_RSOL75 and the M2e3\_E1\_D8e25\_RSOL300 tests is $\approx 2 \%$, while between the M2e3\_E10\_D8e25\_RSOL75 and M2e3\_E10\_D8e25\_RSOL300 tests it is $\approx 3 \%$. The RSOL1200 tests did not have as large of a variation in the final SFE from the RSOL300 tests, and only had a  difference in the SFE of $ \sim 0.35 \%$  and $ \sim 1.2 \%$ for the M2e3\_E1\_D8e25\_RSOL1200 and M2e3\_E10\_D8e25\_RSOL1200 tests respectively.  Since these differences are less than  $1 \sigma$ and we focus this study on the SFE rather than the IMF, we conclude that using $\tilde{c} = 300$ km/s is sufficiently resolved for our simulations. However, the trends in the SFE and stellar mass distribution suggest that future numerical simulations of CR propagation in MCs, particularly studies exploring a high CR environment, should carefully consider the choice of $\tilde{c}$.  

\begin{table*} 
\begin{tabular}{ |c c c| } \hline
 Simulation Name & RSOL [km/s] & SFE (\%) 
 \\
 \hline
M2e3\_E1\_D8e25\_RSOL75 & 75 & 9.3 \\
M2e3\_E1\_D8e25\_RSOL300 & 300 & 7.46  \\
M2e3\_E1\_D8e25\_RSOL1200 & 1200 & 7.09 \\
M2e3\_E10\_D8e25\_RSOL75 & 75 & 10.9  \\
M2e3\_E10\_D8e25\_RSOL300 & 300 & 7.8 \\
M2e3\_E10\_D8e25\_RSOL1200 & 1200 & 6.62  \\
 \hline
\end{tabular}
\caption{Simulation name, RSOL, and final values for the SFE for six different tests varying $\tilde{c}$. All simulations use $N_{\rm gas} = 2 \times 10^5$ and $D_{\parallel} = 8.33 \times 10^{25} \rm cm^2/s$. }
\label{table:table_appendix}
\end{table*}

\section{Tests With Varying Physics}
\label{section:nost_appendix}

Figure \ref{fig:timescale_quantities} shows that for our \fiducial\ simulation both CR transport and energy losses are dominated by streaming, especially at the cloud boundary. As we discuss in Section \ref{subsubsection:cr_exclusion}, these effects cause CRs that originated in the cloud to lose most of their energy and prevent CRs outside the cloud from propagating in without significant energy losses (Figure \ref{fig:discussion_slice_plots}). In order to confirm these conclusions, we run low resolution simulations ($N_{\rm gas}= 2 \times 10^5$) out to $\sim 1.5 t_{\rm ff}$ where we vary the CR transport and loss physics. The first one has identical transport and loss physics to our high resolution simulations. In the second we disable all heating/cooling interactions with the gas, and consequently all collisional losses (Equations \ref{eq:cr_energy_loss} and \ref{eq:lambda_cat_loss}). In the final test simulation we disable both streaming transport and energy losses (all terms including $\vst$ in Equations \ref{eq:cr_energy_one} and \ref{eq:cr_energy_two}). All three of these simulations start with the initial CR background and diffusion coefficient used in the \fiducial\ simulation ($\epscr, _{ \rm med}$ = 1 $\rm eV/cm^3$ and $D_{\parallel} = 8.33 \times 10^{25} \rm cm^2/s$) and use an RSOL of $\tilde{c}=300 \rm km/s$. Since these tests are designed as a simple test of the relevant CR physics and losses, we only run them out to $\sim 1.5 t_{\rm ff}$. The three different runs are listed in Table \ref{table:table_appendix_crphys}. 

\begin{table*} 
\begin{tabular}{ |c c c| } \hline
 Simulation Name & Heating/Cooling? & Streaming?  
 \\
 \hline
M2e3\_E1\_D8e25\_ALL & Y & Y \\
M2e3\_E1\_D8e25\_NOLOSS & N & Y  \\
M2e3\_E1\_D8e25\_NOST & Y & N \\
 \hline
\end{tabular}
\caption{Parameters for different tests varying CR transport and loss physics. All simulations use $N_{\rm gas} = 2 \times 10^5$, $\epscr, _{ \rm med}$ = 1 $\rm eV/cm^3$, and $D_{\parallel} = 8.33 \times 10^{25} \rm cm^2/s$. }
\label{table:table_appendix_crphys}
\end{table*}

\begin{figure*}
\centering
\includegraphics[width=0.99 \textwidth]
{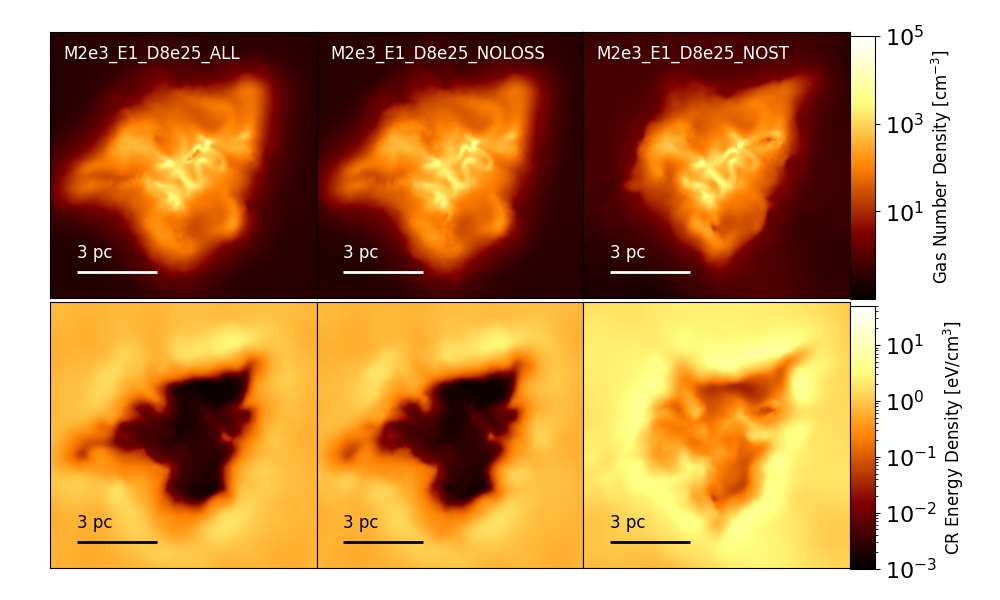}
\caption{Gas number density (top) and CR energy density (bottom) at $1\tff$ for simulations comparing our \fiducial\ setup (left) to a setup with no collisional losses (middle) and a setup with no streaming transport and energy losses (right).}
\label{fig:nost_slice_plots}
\end{figure*}

\begin{figure}[tbh!]
\centering
\includegraphics[width=0.49 \textwidth]
{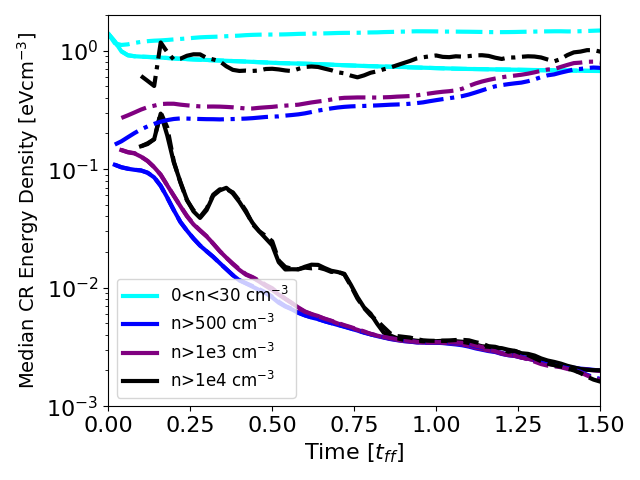}
\caption{Median CR energy density for gas in the indicated density range or lower limit for the M2e3\_E1\_D8e25\_ALL (solid) and M2e3\_E1\_D8e25\_NOLOSS (dashed) and  M2e3\_E1\_D8e25\_NOST (dashdot) simulations.}
\label{fig:nost_evolution_plots}
\end{figure}

Figure \ref{fig:nost_slice_plots} shows slice plots of the gas number density (top) and CR energy density (bottom) of the three different simulations at $t=t_{\rm ff}$. Both the gas number density and CR energy density appear almost identical for the M2e3\_E1\_D8e25\_ALL (left) and M2e3\_E1\_D8e25\_NOLOSS (middle) simulations. Figure \ref{fig:nost_evolution_plots} shows the time evolution of the CR energy density for the three different simulations. The evolution of the median CR energy density is indistinguishable for the M2e3\_E1\_D8e25\_ALL (solid) and M2e3\_E1\_D8e25\_NOLOSS (dashed) simulations. These results confirm our conclusion that CR collisional losses are insignificant compared to streaming energy losses in our simulations. 

In contrast, our simulation with no streaming transport or losses shows significant differences. The right column of Figure \ref{fig:nost_slice_plots} shows that the gas density for the M2e3\_E1\_D8e25\_NOST simulation appears similar to the other two simulations at $t=t_{\rm ff}$ but not indistinguishable. However, the CR energy density (bottom) is much higher than in the other two simulations. This is especially true in the cloud and at the cloud boundary, where it is up to two orders of magnitude higher. Although there is still a slight enhancement in the CR energy density at the cloud boundary due to magnetic field effects (Section \ref{subsubsection:cr_exclusion}), CRs no longer lose energy in this region and can propagate through into the cloud. Figure \ref{fig:nost_evolution_plots} shows that the median CR energy density for this simulation (dashdot) rises at all gas densities inside the cloud. The CRs are advected with the gas as the cloud collapses, leading to the enhancement at higher densities. Finally, the slight increase the CR energy density outside the cloud (cyan) is due to diffusive reacceleration, allowing the CRs to gain energy as they propagate. 

\bibliography{main}{}
\bibliographystyle{aasjournal}

\end{document}